\documentclass[a4paper, draftcls, 12pt, onecolumn, twoside]{IEEEtran}
\IEEEoverridecommandlockouts

\usepackage{graphicx,cite,acronym,amsmath,amssymb,amsfonts,color,subfigure,floatflt,stfloats}
\usepackage{epsfig,epstopdf,psfrag}
\usepackage[table]{xcolor}
\usepackage{bm}

\usepackage[cmintegrals]{newtxmath}
\usepackage{graphics,hhline,array,cite,bbm}

\usepackage{latexsym}
\usepackage{theorem}
\usepackage{times}
\usepackage{makeidx}

\DeclareGraphicsExtensions{.png,.jpg,.eps}

\acrodef{CR}{channel response}

\acrodef{BS}{base station}

\acrodef{MS}{mobile station}

\acrodef{UE}{user equipment}

\acrodef{MIMO}{multiple-input multiple-output}

\acrodef{RIS}{reconfigurable intelligent surface}

\acrodef{LIS}{large intelligent surface}

\acrodef{MIS}{medium intelligent surface}

\acrodef{SIS}{small intelligent surface}

\acrodef{DoF}{degrees-of-fredom}

\acrodef{AF}{amplify \& forward}

\acrodef{DF}{detect \& forward}

\acrodef{JF}{just forward}

\acrodef{CSI}{channel state information}

\acrodef{RV}{random variable}

\acrodef{i.i.d.}{independent, identically distributed}

\acrodef{PSD}{power spectral density}

\acrodef{PDF}{probability distribution function}

\acrodef{CDF}{cumulative distribution function}

\acrodef{ch.f.}{characteristic function}

\acrodef{AWGN}{additive white Gaussian noise}

\acrodef{RSSI}{received signal strength indicator}

\acrodef{SNR}{signal-to-noise ratio}

\acrodef{LRT}{likelihood ratio test}

\acrodef{GLRT}{generalized likelihood ratio test}

\acrodef{GML}{generalized maximum likelihood}

\acrodef{LOS}{line-of-sight}

\acrodef{NLOS}{non-line-of-sight}

\acrodef{GDOP}{geometric dilution of precision}

\acrodef{GPS}{Global Positioning System}

\acrodef{FIM}{Fisher information matrix}

\acrodef{PEB}{position error bound}

\acrodef{WSN}{Wireless Sensor Network}

\acrodef{MAC}{medium access control}

\acrodef{RSS}{received signal strength}

\acrodef{RTT}{round-trip time}

\acrodef{MIMO}{multiple-input multiple-output}

\acrodef{MF}{matched filter}

\acrodef{ED}{energy detector}

\acrodef{ML}{maximum likelihood}

\acrodef{NL}{nonlinear}

\acrodef{MSE}{mean square error}

\acrodef{RMSE}{root mean square error}

\acrodef{ppm}{part-per-million}

\acrodef{PRP}{pulse repetition period}

\acrodef{ACK}{acknowledge}

\acrodef{UWB}{ultrawide bandwidth}

\acrodef{TNR}{threshold-to-noise ratio}

\acrodef{NLOS}{non line-of-sight}

\acrodef{LOS}{line-of-sight}

\acrodef{LS}{least squares}

\acrodef{IR-UWB}{impulse radio UWB}

\acrodef{FCC}{Federal Communications Commission}

\acrodef{TH}{time-hopping}

\acrodef{PPM}{pulse position modulation}

\acrodef{PAM}{pulse amplitude modulation}

\acrodef{MUI}{multi-user interference}

\acrodef{PDP}{power delay profile}

\acrodef{PPP}{Poisson point process}

\acrodef{DS}{delay spread}

\acrodef{CED}{channel excess delay}

\acrodef{BPZF}{band-pass zonal filter}

\acrodef{SIR}{signal-to-interference ratio}

\acrodef{RFID}{radio frequency identification}

\acrodef{WPAN}{wireless personal area networks}

\acrodef{WWLB}{Weiss-Weinstein lower bound}

\acrodef{DP}{direct path}

\acrodef{MF}{matched filter}

\acrodef{MMSE}{minimum-mean-square-error}

\acrodef{SBS}{serial backward search}

\acrodef{NBI}{narrowband interference}

\acrodef{WBI}{wideband interference}

\acrodef{INR}{interference-to-noise ratio}

\acrodef{CIR}{channel impulse response}

\acrodef{ISI}{inter-symbol interference}

\acrodef{CPR}{channel pulse response}

\acrodef{LRT}{likelihood ratio test}

\acrodef{RADAR}{RADAR}

\acrodef{MUR}{Multistatic RADAR}

\acrodef{MUI}{multi-user interference}

\acrodef{e.m.}{electromagnetic}

\acrodef{CW}{continuous wave}

\acrodef{RF}{radiofrequency}

\acrodef{FCC}{Federal Communications Commission}

\acrodef{EIRP}{effective radiated isotropic power}

\acrodef{RCS}{radar cross section}

\acrodef{BAV}{balanced antipodal Vivaldi}

\acrodef{PRake}{partial Rake}

\acrodef{RTLS}{real time locating system}

\acrodef{CRB}{Cram\'{e}r-Rao bound}

\acrodef{ZZB}{Ziv-Zakai bound}

\acrodef{TOA}{time-of-arrival}

\acrodef{TOF}{time-of-flight}

\acrodef{WSN}{wireless sensor network}

\acrodef{MAC}{medium access control}

\acrodef{RSS}{received signal strength}

\acrodef{TDOA}{time difference-of-arrival}

\acrodef{RF}{radiofrequency}

\acrodef{RTT}{round-trip time}

\acrodef{AOA}{angle-of-arrival}

\acrodef{MF}{matched filter}

\acrodef{ED}{energy detector}

\acrodef{ML}{maximum likelihood}

\acrodef{MUR}{Multistatic radar}

\acrodef{HDSA}{high-definition situation-aware}

\acrodef{RRC}{root raised cosine}

\acrodef{OFDM}{orthogonal frequency division multiplexing}

\acrodef{IF}{intermediate frequency}

\acrodef{PHY}{physical layer}

\acrodef{S-V}{Saleh-Valenzuela}

\acrodef{UHF}{ultra-high frequency}

\acrodef{PR}{pseudo-random}

\acrodef{SoC}{System on Chip}

\acrodef{SoP}{System on Package}

\acrodef{SPMF}{Single-Path Matched Filter}

\acrodef{IMF}{Ideal Matched Filter}

\acrodef{SCR}{signal-to-clutter ratio}

\acrodef{BEP}{bit error probability}

\acrodef{BER}{bit error rate}

\acrodef{WSR}{wireless sensor radar}

\acrodef{HPBW}{half power beam width}

\acrodef{LEO}{localization error outage}

\acrodef{WSS}{wide-sense stationary}

\acrodef{TR}{time-reversal}

\acrodef{WSSUS}{WSS with uncorrelated scattering}

\acrodef{GP}{Gaussian process}

\acrodef{IMU}{inertial measurement unit}

\newcommand{\bolds} {{\bf{s}}}

\newcommand{\boldk} {{\bf{k}}}

\newcommand{\boldu} {{\bf{u}}}

\newcommand{\boldX} {{\bf{X}}}

\newcommand{\boldH} {{\bf{H}}}

\newcommand{\boldI} {{\bf{I}}}
\newcommand{\boldr} {{\bf{r}}}

\newcommand{\SNR}{\text{SNR}}

\newcommand{\boldsp} {{\bf{s{'}}}}
\newcommand{\boldrp} {{\bf{r{'}}}}

\newcommand{\versorr} {\hat{\bf{r}}}
\newcommand{\versorp} {\hat{\bf{p}}}

\newcommand{\versorn} {\hat{\bf{n}}}

\newcommand{\versorx} {{\hat{\bf{u}}_x}}
\newcommand{\versory} {{\hat{\bf{u}}_y}}
\newcommand{\versorz} {{\hat{\bf{u}}_z}}

\newcommand{\rx} {r_x}
\newcommand{\ry} {r_y}

\newcommand{\sx} {s_x}
\newcommand{\sy} {s_y}
\newcommand{\sz} {s_z}

\newcommand{\xo} {x_{0}}
\newcommand{\yo} {y_{0}}

\newcommand{\hatbz} {{\bf{\hat{z}}}}

\newcommand{\hatbn} {{\bf{\hat{n}}}}

\newcommand{\Vt} {V_{\text{T}}}
\newcommand{\Vr} {V_{\text{R}}}
\newcommand{\Sr} {\mathcal{S}_{\text{R}}}
\newcommand{\St} {\mathcal{S}_{\text{T}}}
\newcommand{\Kt} {K_{\text{T}}}
\newcommand{\Kr} {K_{\text{R}}}

\newcommand{\Sx} {S_{\text{x}}}
\newcommand{\Sy} {S_{\text{y}}}
\newcommand{\Lx} {L_{\text{x}}}
\newcommand{\Ly} {L_{\text{y}}}
\newcommand{\Lz} {L_{\text{z}}}
\newcommand{\At} {A_{\text{T}}}
\newcommand{\Ar} {A_{\text{R}}}
\newcommand{\AR} {{\text{AR}}}

\newcommand{\Gt} {G_{\text{T}}}
\newcommand{\Gr} {G_{\text{R}}}
\newcommand{\Gi} {G_{\text{I}}}

\newcommand{\Green} {{\bf \it G} }
\newcommand{\Em} {{\bf \it E}}
\newcommand{\Jm} {{\bf \it J}}

\newcommand{\um} {{\bf \it u}}
\newcommand{\vm} {{\bf \it v}}

\newcommand{\bphi} {\it \phi}
\newcommand{\bpsi} {\it  \psi}

\newcommand{\Upx} { U_x^{+}(r_x) }
\newcommand{\Umx} { U_x^{-}(r_x) }
\newcommand{\Upy} { U_y^{+}(r_y) }
\newcommand{\Umy} { U_y^{-}(r_y) }

\begin{document}
\title{Communicating with Large Intelligent Surfaces: Fundamental Limits and Models}

\author{
\IEEEauthorblockN{Davide~Dardari,~\IEEEmembership{Senior~Member,~IEEE}}
\IEEEcompsocitemizethanks{\IEEEcompsocthanksitem 
 D.~Dardari is with the 
   Dipartimento di Ingegneria dell'Energia Elettrica e dell'Informazione ``Guglielmo Marconi"  (DEI), CNIT, 
   University of Bologna, Cesena Campus, 
   Cesena (FC), Italy, (e-mail: davide.dardari@unibo.it). 
    }
}

\maketitle
\begin{abstract}

This paper analyzes the optimal communication involving large intelligent surfaces (LIS) starting from electromagnetic arguments.
Since the numerical solution of the corresponding  eigenfunctions problem is in general computationally prohibitive, simple but accurate analytical expressions for the link gain and available spatial degrees-of-freedom (DoF) are derived.   
It is shown that the achievable DoF and gain offered by the wireless link are determined only by geometric factors, and that the classical Friis' formula is no longer valid in this scenario where the transmitter and receiver could operate in the near-field regime.
Furthermore, results indicate that, contrarily to classical  MIMO systems, when using LIS-based antennas   DoF larger than 1 can be exploited even in strong line-of-sight  (LOS) channel conditions, which corresponds to a significant increase in spatial capacity density, especially when working at millimeter waves.
\end{abstract}


\section{Introduction}

\IEEEPARstart{F}{uture}  wireless networks are expected to become distributed intelligent communication, sensing and computing entities. This will allow to meet ultra-reliability, high capacity densities, extremely low-latency and low-energy consumption requirements posed by emerging application scenarios such as  Industrial Internet of Things in Factories of the Future \cite{VitZunSau:19,Zha:19}.
The current trend in satisfying part of such requirements is through cell densification, massive \ac{MIMO} transmission, and the exploitation of higher frequency bands (e.g., millimeter and  THz) \cite{San:19,RapXinKanJuMadManAlkTri:19}. 
Unfortunately, when moving to higher frequency bands the channel path-loss increases and the multipath becomes sparse so that the spatial multiplexing peculiarity of \ac{MIMO}, i.e., the channel \ac{DoF}, guaranteed at lower frequencies by rich multipath, is lost in favor of only beamforming gain which increases the communication capacity logarithmically instead of linearly with the number of antennas \cite{TseVis:B05}. 

The introduction of  metamaterials  to realize, for instance, the so called \emph{metasurfaces} \cite{HolKueGorOHaBooSmi:12,Gly:16,Bur:16},   has attracted a wide interest in different research communities with  applications including  transmitarrays  \cite{DiP:17}, metamirrors \cite{RadAsaTre:14, AsaAlbTcvRubRadTre:16,AsaRadVehTre:15}, reflectarrays 
\cite{HumCar:14,NayYanEsh:15}, metaprisms \cite{DarMas:20}, and holograms \cite{Gly:16,Hun:14}. 
Furthermore, the recent development of programmable metasurfaces, used as  
smart \ac{e.m.} reflectors and large configurable antennas, has opened new very appealing perspectives \cite{Sil:14,DiP:17,NepBuf:17,GonMinChaMac:2017,Eldar:19}.
In fact, these \emph{intelligent surfaces} can be easily embedded in daily life objects such as walls, clothes, buildings, etc..   
Environments coated with intelligent surfaces constitute the recently proposed \emph{smart radio environments} concept \cite{LiaNieTsiPitIoaAky:18,DiRenzo:19,DiRenzo:19b}. In smart radio environments, the design paradigm is changed from wireless devices/networks that adapt themselves to the environment (e.g., propagation conditions), to the joint optimization of both devices and environment using \acp{RIS}. 

The advantages of \ac{RIS}-enabled systems have been analyzed in several papers.  For instance, in \cite{YanZhaZha:20} a \ac{RIS}-enhanced \ac{OFDM} system is investigated, where the power allocation and the phase profile of the metasurface  are jointly optimized to boost the achievable rate of a cell-edge user. 
In \cite{ZhaZha:19,OzdBjoLar:20}, it is shown that  the channel rank of  \ac{MIMO} communication in \ac{LOS}  can be increased by adding a \ac{RIS} generating an artificial path that can be exploited by the \ac{MIMO} system to increase the capacity. The authors in \cite{DiRenzo:19c} present a comparison between \ac{RIS}- and relay-enabled wireless networks by discussing the similarities and differences. Other studies can be found, for instance, in \cite{YeGuoAlo:19,JunSaaDebHon:19}. An interesting alternative to \acp{RIS} is given by \emph{metaprisms}, which are passive and non-configurable frequency-selective metasurfaces proposed in \cite{DarMas:20}. With an appropriate design of the metaprism, it is possible to control that each data stream in an \ac{OFDM} system is reflected to the desired direction by properly dispatching subcarriers to users. This helps to cover areas experiencing severe \acf{NLOS} channel conditions at low-cost.  

\begin{figure}[t]
\centerline{\includegraphics[width=0.8\columnwidth]{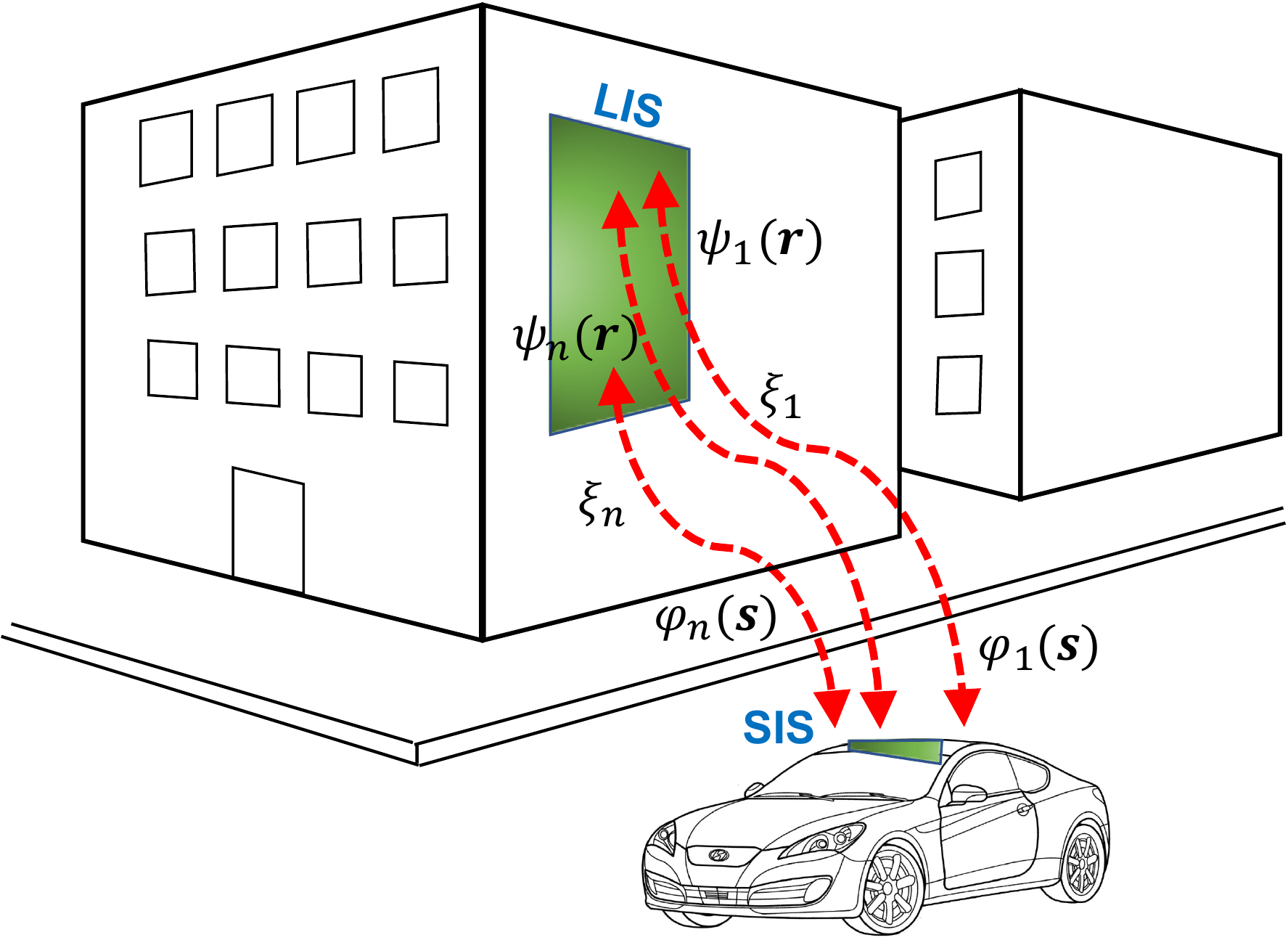}}
\caption{Example of LIS-based communication scenario.}
\label{Fig:Scenario}
\end{figure}

\subsection{Related Work}
Most of the papers dealing with intelligent surfaces, such as those cited above, use them as (possibly reconfigurable) reflectors to assist multipath propagation.  
A few papers analyze the potential of using metasurfaces as large antennas, namely \ac{LIS}-based antennas, to improve the communication capacity \cite{HuRusEdf:18,NepBuf:17,Eldar:19} or to enable single-anchor localization \cite{GuiDar:19}.  

One question is whether smart metasurfaces can be used as the enabling technology to approach the full control of \ac{e.m.} waves generated or sensed by antennas. In fact, with metamaterials \ac{e.m.} waves can be shaped almost arbitrarily, at least in theory. It is expected that this unprecedented flexibility offered by \acp{LIS} (e.g., on walls), or \acp{MIS} (e.g., on cars/truck), and \acp{SIS} (e.g., on smartphones/sensors), will provide a great opportunity to move  towards the ultimate capacity limit of the wireless channel. 

The main fundamental results on the physical limitation brought by the \ac{e.m.} transportation of information can be found in \cite{BucFra:89,FraMigDonMin:09,FraMigMinSch:11,FranceschettiBook:2018} and references therein. 
The authors in \cite{PizMarSan:19b}  extend these results by generalizing the Clarke's channel model to non-isotropic random scattering environments using a mathematically tractable framework based on Fourier plane-wave series expansion of the channel. With this model they show that the \ac{DoF} of the \ac{e.m.} wave on a continuous large antenna aperture is proportional to the surface area normalized to the square wavelength \cite{PizMarSan:19}.
These results mainly address the computation of the spatial dimensionality of the \ac{e.m.} field when considering finite volumes with sources and scatterers in the far-field, but they do not consider the spatial \ac{DoF} available in a communication system employing intelligent surfaces  as transmit and receive antennas, possibly in the near-field region.

The adoption of \ac{LIS}-enabled antennas provides high  flexibility  in network design as well as the potential to achieve the goals of next generation wireless networks, but it has also opened several fundamental questions that are still unsolved, such as understanding the theoretical limits  and how to achieve them in practice.

With \acp{LIS}, classical models for antenna arrays fail to capture the actual wireless link characteristics in terms of gain, path-loss and available \ac{DoF} as they assume  (Fraunhofer) far-field condition (i.e., a distance much larger than the antenna dimension so that waves can be considered plane \cite{HarringtonBook:2001}), whereas with \acp{LIS} the size of the antenna becomes comparable to the distance of the link (near-field regime). Moreover, they usually do not account for the flexibility in generating the current distribution offered by \acp{LIS} (holographic capability) and hence common results of aperture antennas are no longer valid. 
Therefore, new models based on the ultimate physical limitation brought by the \ac{e.m.} transportation of information should be considered.

One of the earliest works proposing and studying \acp{LIS} for communication is \cite{HuRusEdf:18}, which considers  the communication between a single-antenna user with a  \ac{LIS}, where  an analysis of the  spatial capacity density   is presented. 
Practical aspects related to the design of the optimal sampling lattice of the  \ac{LIS}  are considered by showing that the
hexagonal lattice is optimal for minimizing the surface's area of a \ac{LIS} under the constraint that one independent signal dimension should be obtained per spent antenna element of the \ac{LIS}.

The work has been extended in \cite{HuChiRusEdf:18} by the same authors to investigate the optimal user assignments for a distributed \ac{LIS} system with several \ac{LIS} units, with the purpose to  select the set of best units to serve a given number of users simultaneously.
The multi-user scenario is considered also in \cite{AmiAngDeCHea:18}, where it is shown that when using a massive \ac{MIMO} system with extremely large arrays, users can effectively communicate only with a sub-part of the array, thus creating non-stationary patterns. The paper proposes a receiver architecture based on subarray processing capable of dealing with this situation.  

In \cite{JunSaaJanKonChgo:19}, the distribution of the sum-rate of an uplink \ac{LIS} network under imperfect channel estimation is investigated through an asymptotic analysis, from which expressions for the outage probability are derived and used to show that \ac{LIS}-based systems  can provide
reliable communications. Further results can be found in the recent  paper \cite{JunSaaJanKoCho:20} published by the same authors, where  the occurrence of channel hardening effects is also analyzed. 

In \cite{Zha:2018} a general theory of space-time modulated digital coding metasurfaces to obtain simultaneous manipulations of \ac{e.m.} waves in both space and frequency domains is proposed and validated in the far-field regime. 

Finally, the issue of power and cost of large massive \ac{MIMO} systems using metasurfaces is addressed in \cite{Eldar:19}. 
Such challenges are tackled by incorporating signal processing methods, such as compression and analog combining, in the physical antenna structure.
The characterization of the maximal achievable sum-rate on the uplink and potential gain over standard antenna arrays are studied.

\subsection{Main Contribution}
To the author's knowledge, no results are present related to the investigation of the available spatial \ac{DoF} as well as the coupling gain between  intelligent surfaces,  in particular when the maximum degree of flexibility in \ac{e.m.} shaping  is allowed, and one of the antennas is large so that it might operate in the near-field even at practical distances.  

In this paper, the optimal communication between \ac{LIS}/\ac{SIS}  is addressed as an eigenfunctions problem starting from an \ac{e.m.} formulation, similarly to what done in \cite{Miller:00,PieMil:00} for optical systems, and preliminarly addressed in \cite{Dar:19b}. 
Unfortunately, finding the solution to the eigenfunctions problem requires extensive and sometimes prohibitive \ac{e.m.}-level simulations if large surfaces are considered, and usually they do not provide general insights. Therefore, we focus on obtaining approximate  but accurate analytical expressions  for the link gain and the available orthogonal communication channels (i.e., the \ac{DoF})  between the transmitter and receiver. 
Although such expressions are easy to compute numerically,  we further derive   closed-form asymptotic and non-asymptotic expressions for some specific cases of interest which allow to get important insights about the communication between intelligent surfaces and can serve as design guidelines in future wireless networks employing \acp{LIS}. 

Furthermore, we show that, contrarily to classical \ac{MIMO},  
 with \ac{LIS}-based antennas the available \ac{DoF} can be higher than 1, even in \ac{LOS} channel condition, 
 thus  boosting in principle the channel capacity. 
 This is useful when moving towards high frequencies where \ac{LOS} communication becomes predominant and the multipath weaker so that conventional \ac{MIMO} systems cannot benefit from multiplexing  gain (i.e., the \ac{DoF}), usually obtained by exploiting the multipath.
In addition, the achievable \ac{DoF} and gain offered by the wireless link are shown to be determined only by geometric factors normalized to the wavelength, and that the classical Friis' formula is no longer valid when using \acp{LIS}. 
Asymptotic expressions for very large \acp{LIS} or large distances put in evidence the difference between classical and \ac{LIS}-based communication systems. 

The remainder of this paper is organized as follows. In
Section~\ref{Sec:Problem}, the general  problem formulation is given.   
Analytical expressions for the link gain and the communication \ac{DoF} for the general case as well as for same particular geometric configurations are derived, respectively, in Sections~\ref{Sec:Power} and \ref{Sec:DoF}.
Numerical results and discussions  are presented in Section~\ref{Sec:NumericalResults}. Finally,
conclusions are given in Section.~\ref{Sec:Conclusion}.

\subsection{Notation and Definitions}
Lowercase bold variables denote vectors in the 3D space, i.e.,  $\boldr=\boldu_x \cdot r_x + \boldu_y \cdot r_y + \boldu_z\cdot r_z$
 is a vector with cartesian coordinates $(r_x, r_y, r_z)$,  $\versorr$ is a unit vector denoting its direction, and  $r=|\boldr|$ denotes its magnitude, where $\versorx$, $\versory$ and $\versorz$ represent the unit vectors in the $x$, $y$ and $z$ directions, respectively.   
Italic capital letters (e.g., $\Em(\boldr)$, $\Jm(\boldr)$) represent electromagnetic vector functions. Boldface capital letters are matrices (e.g., $\boldH$), where $\boldI$ is the identity matrix, and $^{\dag}$ indicates the conjugate transpose operator.
$\nabla^2  \, \Jm (\boldr)$ is the Laplacian of the vector function $\Jm(\boldr)$,  whereas $\nabla \Phi$ and $\nabla \cdot \Jm (\boldr)$ are the gradient and divergence operators, respectively.  
Surfaces and volumes are indicated with calligraphic letters $\St$, where $\At=|\St|$ is their Lebesgue measure.
Define the ${\cal{L}}_2$-norm $||\boldr ||$, the Frobenius norm 
$|| \boldX ||=\sqrt{\sum_{k=1}^N \sum_{j=1}^N \left | \left \{ \boldX \right \}_{kj} \right |^2}$,  
and the outer product (tensor product)  $\boldr \otimes \bolds$, where $\{ \boldr \otimes \bolds \}_{kj}=r_k \, s_j$, and $\left \{ \boldX \right \}_{kj}$ is the $kj$th element of matrix $\boldX$.
 The notation ${\mathbb{L}}^2(\St)$ indicates the Hilbert space corresponding to the square-integrable functions defined on $\St$. 
Furthermore, denote with $\mu$, $\epsilon$, and $\eta=\sqrt{\mu/\epsilon}$ the  permittivity, permeability and impedance of free-space, respectively, and $c$  the speed-of-light.

\section{General Problem Formulation}
\label{Sec:Problem}

Thanks to the adoption of metamaterials, with \acp{LIS}  one can synthesize in principle any current distribution, then it is of interest to investigate how many orthogonal channels, i.e.,  \ac{DoF},  can be established when two \ac{LIS}/{SIS} are communicating with each other. 
To this purpose, we approximate the intelligent surface as a continuous array of an infinite number of infinitesimal antennas.  A system having an uncountably infinite number of antennas in a finite space has been recently dubbed as \emph{Holographic \ac{MIMO}} \cite{PizMarSan:19b}. 

\subsection{Problem Formulation}

Consider a transmit  \ac{LIS} or \ac{MIS}/\ac{SIS} antenna with surface $\St$ of area $\At=|\St|$ containing \ac{e.m.} monochromatic source currents  with Fourier representation $\Jm(\bolds\, ,\omega)$ different from zero in   $\bolds \in \St$, with $\omega$ being the angular frequency, which generate an electric field  $\Em(\boldr\, ,\omega)$  at the generic location $\boldr$ in free-space. Furthermore, we consider  a receive \ac{LIS} antenna $\Sr$ not intersecting $\St$, with area $\Ar=|\Sr|$.\footnote{Here we consider only surfaces because of their higher practical relevance, even though most of the following results can be extended to volumes as well.}  Due to the reciprocity of the radio medium, their role can be exchanged.

Each frequency component satisfies the  inhomogeneous Helmholtz  wave equation\footnote{Similar formulation can be done for the magnetic field in case of magnetic currents even though it is always possible to model the problem using equivalent source currents \cite{BalB:16}.}
\begin{equation} \label{eq:waveeq}
 \nabla ^2 \Em(\boldr) + k_0^2 \, \Em(\boldr) =  \jmath k_0 \, \eta\,  \Jm(\boldr) \, ,
\end{equation}
where $k_0=\omega/c=2\pi/\lambda$ is the wavenumber, $\lambda$ the wavelength, and we have dropped the explicit dependence on $\omega$ to lighten the notation.  

Any point source in $\St$ generates the (outgoing) wave given by the  tensor Green's function \cite{HarringtonBook:2001}  
\begin{equation}  \label{eq:Green1}
 \Green(\boldr)=-\frac{\jmath \omega \mu }{4\pi } \left [\boldI + \frac{1}{k_0} \nabla \nabla   \right ] \frac{\exp \left ( -\jmath k_0  r   \right )}{ r} \, ,
 \end{equation}
with $r=|\boldr |$, which obeys the Helmholtz equation. 

By expanding \eqref{eq:Green1} we obtain  \cite{PooBroTse:05}
 \begin{align}   \label{eq:Green2}
   \Green(\boldr )=& - \frac{\jmath \, \eta \exp \left ( -\jmath k_0 r   \right )}{2\lambda r } \left [ \left (\boldI - \versorr \cdot \versorr^{\dag}  \right ) 
 + \frac{\jmath \lambda}{2\pi r} \left (\boldI - 3\, \versorr \cdot \versorr^{\dag}  \right ) \right .  \\
 & \left . - \frac{ \lambda^2}{(2\pi r)^2} \left ( \boldI - 3\, \versorr \cdot \versorr^{\dag}  \right )\right ] 
 \simeq - \frac{\jmath \, \eta \exp \left ( -\jmath k_0 r   \right )}{2\lambda r } \left (\boldI - \versorr \cdot \versorr^{\dag}  \right ) \, ,  \nonumber  
 \end{align}
where we grouped the terms multiplying, respectively, the factors $1/r$, $1/r^2$ and $1/r^3$. 
It is evident from \eqref{eq:Green2} that when $r\gg \lambda$, the second and third terms can be neglected and hence the right-hand side approximation in \eqref{eq:Green2} holds.\footnote{Under this condition the system does not work in the  `reactive' near-field.}  
By adding all the waves from the sources in $\St$, the resulting wave in $\boldr$ is
\begin{equation} \label{eq:GreenOperator}
 \Em(\boldr)=\int_{\St} \Green(\boldr - \bolds)\, \Jm(\bolds) \, d\bolds  \, .
 \end{equation}

The goal is to determine how many orthogonal communication channels,  namely \emph{communication modes}, with as large coupling intensities as possible can be activated between $\St$ and $\Sr$. 
This is associated to the optimal approximation of every element in the image space of a Hilbert-Schmidt operator in terms of singular functions. 
Specifically, define $\mathcal{X}={\mathbb{L}}^2(\St)$ and $\mathcal{Y}={\mathbb{L}}^2(\Sr)$ the Hilbert spaces corresponding to the square-integrable functions defined in $\St$ and $\Sr$, respectively. The function $\Em(\boldr) \in \mathcal{Y}$ can be seen as the image of $\Jm(\bolds) \in \mathcal{X}$ through the Hilbert-Schmidt kernel $\Green(\boldr,\bolds)=\Green(\boldr - \bolds)$ on $\St \times \Sr$, which induces the operator $\Green:\mathcal{X} \rightarrow \mathcal{Y}$ such that, for any $\Jm \in \mathcal{X}$, 
\begin{equation}
\left ( \Green\, \Jm   \right ) (\boldr)=\int_{\St} \Green(\boldr,\bolds) \, \Jm(\bolds)\, d\bolds \, . 
\end{equation}

Define the following self-adjoint Hilbert-Schmidt operators $\Green^{\dag} \Green$ and $\Green \Green^{\dag}$, with symmetric kernels
\begin{align} 
  \Kt(\bolds\, , \boldsp)&=\int_{\Sr} G^{\dag}(\boldr- \bolds) \, G(\boldr- \boldsp)  \, d \boldr  \\
  \Kr(\boldr\, , \boldrp)  &= \int_{\St} G(\boldr - \bolds) \, G^{\dag}(\boldrp- \bolds)  \, d \bolds \, .
\end{align}

A fundamental property of Hilbert-Schmidt operators is that they are compact and admit either a finite or countably infinite orthonormal basis.
In particular,  
two  sets of orthonormal eigenfunctions $\{\bphi_n(\boldr) \}$, $\{\bpsi_n(\bolds)\}$ exist,  which are solutions, respectively, of the following coupled eigenfunction problems: 
\begin{align} \label{eq:xi1}
 \xi^2 \Jm(\bolds) &= \int_{\St} \Kt(\bolds\, , \boldsp) \, \Jm(\boldsp) \, d \boldsp   \\ \label{eq:xi2}
  \xi^2 \Em(\boldr) &= \int_{\Sr} \Kr(\boldr\, , \boldrp) \, \Em(\boldrp) \, d \boldrp  \, ,
\end{align}
with the same real eigenvalues $\xi_1^2\ge \xi_2^2 \ge \xi_3^2 \ldots$ \cite{FranceschettiBook:2018,Miller:19}.
Note that $\{ \bphi_n(\boldr) \}$ and $\{ \bpsi_n(\boldr) \}$ are two sets of orthonormal (vector) functions that are complete, respectively, in $\St$ and $\Sr$, i.e., 
\begin{align} \label{eq:orthogonal}
\int_{\St} \bphi_n(\boldr) \, \bphi_m^{\dag}(\boldr) \, d\boldr &=\delta_{nm}  &
\, \, \, \int_{\Sr} \bpsi_n(\boldr) \, \bpsi_m^{\dag}(\boldr) \, d\boldr  =\delta_{nm} \, ,
\end{align}
being $\delta_{nm}$ the Kronecker delta.

As a consequence, any current density and wave in $\St$ and $\Sr$ can be written, respectively,  as 
\begin{align} \label{eq:JmEm}
 \Jm(\boldr)& =\sum_n a_n \, \bphi_n(\boldr)  
 & \Em(\boldr) =\sum_n b_n \, \bpsi_n(\boldr) \, ,
\end{align}
being $a_n$ and $b_n$ the inner products, respectively, of $\Jm(\boldr)$ and $\bphi_n(\boldr)$, and  of $\Em(\boldr)$ and $\bpsi_n(\boldr)$. It can be easily verified that $b_n=\xi_n \, a_n$.
 
Consider now the following approximation of the kernel $\Green$ in terms of $D$ singular functions
\begin{align} \label{eq:BilinearExpansion}
\Green_D(\boldr,\bolds)=\sum_{n=1}^{D} \lambda_n \, \um_n(\boldr) \otimes \vm_n^{\dag}(\bolds)  \, .
\end{align}

For a fixed $D$, the best approximation of $\Green(\boldr,\bolds)$ is obtained by choosing in \eqref{eq:BilinearExpansion} $\um_n(\boldr)=\psi_n(\boldr)$, $\vm_n(\bolds)=\phi_n(\bolds)$, and $\lambda_n=\xi_n$, $n=1,2,\ldots D$, so that the error 
\begin{equation}
e_D=\int_{\St} \int_{\Sr} || \Green(\boldr,\bolds)-\Green_D(\boldr,\bolds)  ||^2 \, d\boldr \, d\bolds=\sum_{n=D+1}^{\infty}  \xi_n^2
\end{equation}
is minimized. This implies an optimal $D$-dimensional approximation of any function in $\mathcal{Y}$, image of the operator induced by the kernel $\Green$. 

Since $\xi_D \rightarrow 0$, it is possible to optimally approximate the current density  and wave, respectively,  in $\St$ and $\Sr$, using \eqref{eq:JmEm} up to the first $D$ terms with an error associated with the approximation at any level of accuracy.

The geometric interpretation of this result is that the generic source current $\Jm(\bolds)$ can be projected onto the coordinate system determined by the orthogonal (vector) eigenfunctions $\{\bphi_n(\bolds)\}$ then, through the kernel (or tensor) $\Green(\boldr)$ in \eqref{eq:GreenOperator}, the $n$th eigenfunction $\bphi_n(\bolds)$ of surface $\St$ is put in one-to-one correspondence with the  $n$th eigenfunction $\bpsi_n(\boldr)$ of the receive surface $\Sr$ through the scaling singular value $\xi_n$.   
Therefore, if one takes as source function the $n$th eigenfunction, i.e., $\Jm(\bolds)=\bphi_n(\bolds)$, $\bolds \in \St$, then the output electric field results $\xi_n \, \bpsi_n(\boldr)$, $\boldr \in \Sr$.

The eigenfunction decomposition ensures that the current distribution $ \bphi_1(\bolds)$ in $\St$ leads to the electric field $\xi_1 \, \bpsi_1(\boldr)$ within $\Sr$ with the largest intensity (eigenvalue $\xi_1^2$). The current distribution $\bphi_2(\bolds)$ in $\St$ leads to the electric field $\xi_2 \, \bpsi_2(\boldr)$ within $\Sr$, orthogonal  to $\xi_1 \, \bpsi_1(\boldr)$, with the second largest intensity (eigenvalue $\xi_2^2$), and so on.
Each pair of functions $\left ( \bphi_n(\bolds),\bpsi_n(\boldr) \right )$  determines a spatial dimension of the system (\emph{communication mode}) across which one can establish an orthogonal communication (see Fig. \ref{Fig:Scenario}).

\begin{figure}[t]
\centerline{\includegraphics[width=1\columnwidth]{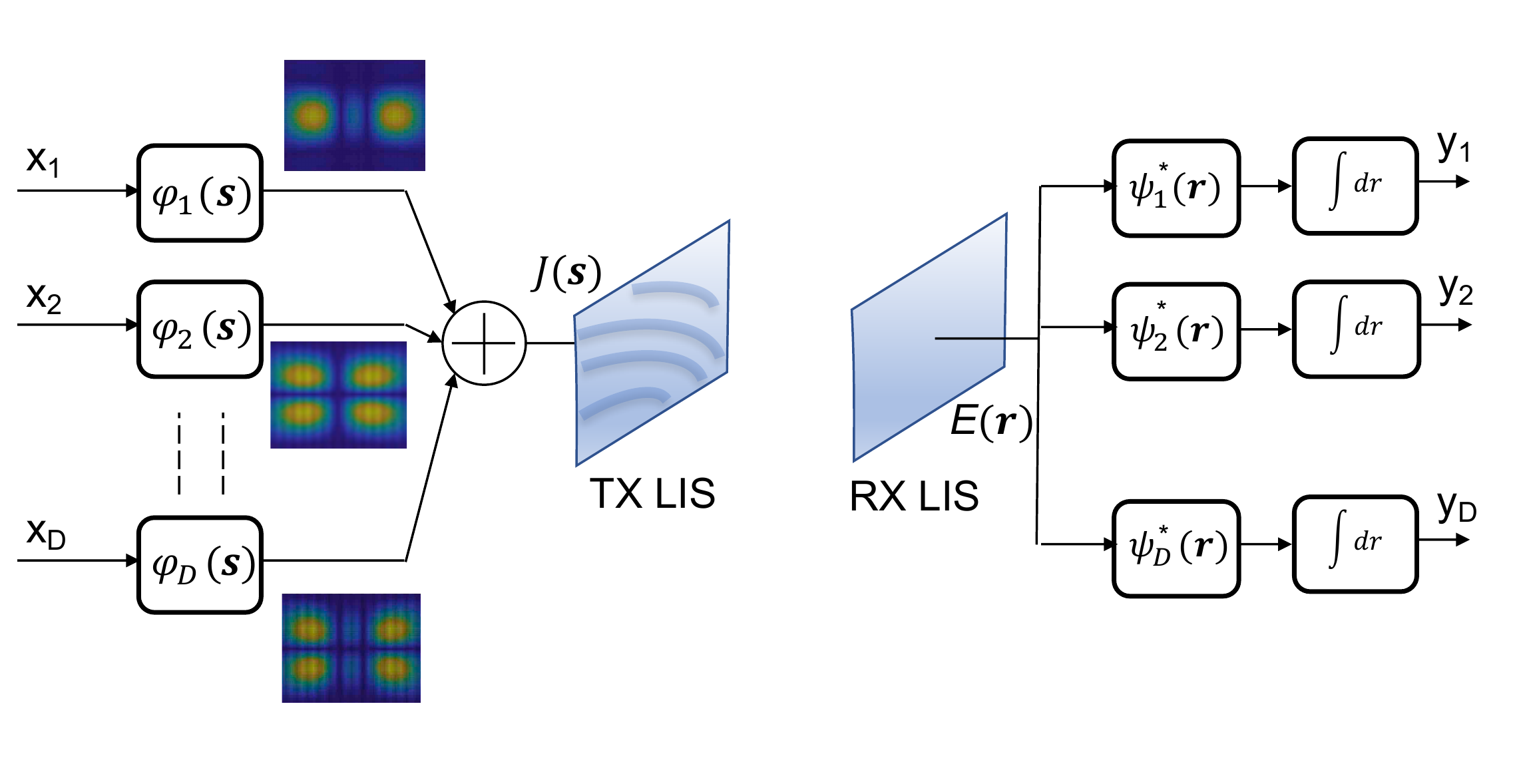}}
\caption{Communication architecture based on orthogonal parallel channels.}
\label{Fig:Parallel}
\end{figure}

It is worth to point out that  since in general the number of eigenvalues in the coupled eigenfunction problems \eqref{eq:xi1} is infinity, the number of communication modes, namely the \ac{DoF}, is defined conventionally as the minimum number $D$ of eigenvalues  sufficient to describe the signals within a given level of accuracy, e.g., compared to the noise intensity. A large level of coupling means that the generated wave is confined approximatively within the space between $\St$ and $\Sr$. Instead, a low level of coupling denotes that the generated wave is mainly dispersed away from the receiver's surface $\Sr$.

By contrast, in \ac{MIMO} systems the  \ac{DoF} corresponds to the rank of the channel matrix which is always no larger  than the minimum between the number of transmit and receive antennas.

In terms of communication system representation, the eigenfunction decomposition leads to the optimal communication architecture depicted in Fig. \ref{Fig:Parallel}.    
From it we obtain the input-output representation in terms of $D$ parallel channels
\begin{align}
y_n=\xi_n \, x_n + w_n \, , \, \, \, \, \, \, \, n=1,2, \ldots , D \, ,
\end{align} 
being $w_n$ the \ac{AWGN}, where the $D$ input data streams $\{ x_n\}$  are associated to the basis functions $\{ \phi_n(s)\}$ in $\St$ (i.e., current spatial distribution on the transmit surface), and they are recovered at the receiver after the  correlation of the received signal $\Em(\boldr)$ with the corresponding   basis functions $\{ \bpsi_n(\bolds) \}$ in $\Sr$. This scheme is  information-theoretical optimal.  

It is worthwhile to highlight that the capacity gain with respect to the case where $D=1$, for a given \ac{SNR}, could be significant. 
For example, supposing uniform power allocation  among the $D$ parallel channels,  such gain  is    
\begin{equation}  \label{eq:capacitygain}
	G_C= \frac{ D \log_2 \left ( 1 +  \frac{\SNR}{D}  \right  )}{\log_2 \left ( 1 +  \SNR  \right ) } \, ,  
\end{equation}
which increases with $D$.

\subsection{Maximum Coupling Intensity Between Intelligent Surfaces}

The effect of each polarization direction can be studied separately if the components of $\Jm(\bolds)=J_x(\bolds) \, \versorx + J_y(\bolds) \, \versory+  J_z(\bolds)\, \versorz$ are taken orthogonal. Therefore, without loss of generality, suppose 
we excite the $x$-component, i.e., $\Jm(\bolds)=J_x(\bolds) \, \versorx$.  
By exploiting the identity \eqref{eq:sumrule} in Appendix A and considering the last term of \eqref{eq:Green2},  the total normalized (i.e., dimensionless) coupling intensity between intelligent surfaces results
\begin{align} \label{eq:coupling2}
 c_x&=\frac{(4\pi)^2}{\lambda^2} \frac{1}{(\omega \mu)^2} \sum_n \xi_n^2= \frac{4}{\eta^2} \int_{\Sr} \int_{\St} ||\Green_x(\boldr -\bolds)||^2   \, d \boldr \, d \bolds \nonumber \\
 &= \frac{1}{\lambda^2} \int_{\Sr} \int_{\St} \frac{(r_y-s_y)^2+(r_z-s_z)^2}{|\boldr - \bolds|^4} \, d \boldr \, d \bolds \, ,
\end{align}
where $\boldr=(r_x,r_y,r_z) \in \Sr$ and $\bolds=(s_x,s_y,s_z) \in \St$ represent the coordinates of the generic points on the receive and transmit surfaces, respectively.

The notation $\Green_x(\cdot)$ indicates we consider only the first column of tensor $\Green(\cdot)$, corresponding to the contribution caused by an excitation in the $x$-direction.
Note that in general the excitation in the $x$-direction might contribute to all directions in the received electric field.
The expressions for the other exciting directions are similars with mutual exchange of $x$, $y$ and $z$.

In \cite{Miller:00} the approximate solution to the eigenfunction problems valid for two collinear rectangular prisms at distance $d$,  oriented along the $z$-axis, of volume $\Vt=\Delta x_{\text{T}} \, \Delta y_{\text{T}}  \, \Delta z_{\text{T}}$ and $\Vr=\Delta x_{\text{R}} \, \Delta y_{\text{R}}  \, \Delta z_{\text{R}}$, respectively, is presented.
Specifically, the solution holds when the volumes are far apart compared to their sizes, i.e., $d \gg  \Delta x_{\text{T}} ,  \Delta y_{\text{T}}  , \Delta z_{\text{T}}, \Delta x_{\text{R}} , \Delta y_{\text{R}}  , \Delta z_{\text{R}}$, which means they are in the (Fraunhofer) far-field region.
In this case, the  \ac{DoF}  available for communication has been found to be 
\begin{equation} \label{eq:Dmiller}
	D=\frac{\Delta x_{\text{T}} \, \Delta y_{\text{T}} \Delta x_{\text{R}} \, \Delta y_{\text{R}} }{d^2 \lambda^2} \, ,
\end{equation}
whereas the total (un-normalized) coupling factor is 
\begin{equation} \label{eq:gainMiller}
  c = \frac{\Vt \, \Vr}{(4 \pi d)^2} \, .
\end{equation} 

Note that the thickness of volumes in the $z$ axis does not  affect the \ac{DoF} but only the coupling intensity.

Incidentally, for very small antennas, i.e., $\Delta x_{\text{T}} \Delta x_{\text{R}} \ll \lambda \, d$,  $\Delta y_{\text{T}} \Delta y_{\text{R}} \ll  \lambda \, d$,  only one solution to the eigenfunction problems exists, corresponding to a plane wave that travels with direction from the transmit antenna to the receive antenna. 
Unfortunately, the result above by \cite{Miller:00} is no longer valid when analyzing a \ac{LIS}  as the assumption of far apart antennas, and hence the parallax approximation typical of the Fraunhofer region, does not hold anymore.

The analytical derivation of the eigenfunctions and eigenvalues is in general elusive and one has to resort to \ac{e.m.} simulations, which could be prohibitive for \acp{LIS} and typically they do not provide general insights.   
In the next sections we bypass the direct derivation of the solutions to the eigenfunction problems by resorting to geometric arguments, with the purpose to determine the spatial \ac{DoF} available for communication. 
Our aim is to derive simple expressions for  particular geometric configurations of interest, also valid  in the radiating near-field.

 \begin{figure}[t]
\centerline{\includegraphics[width=0.8\columnwidth]{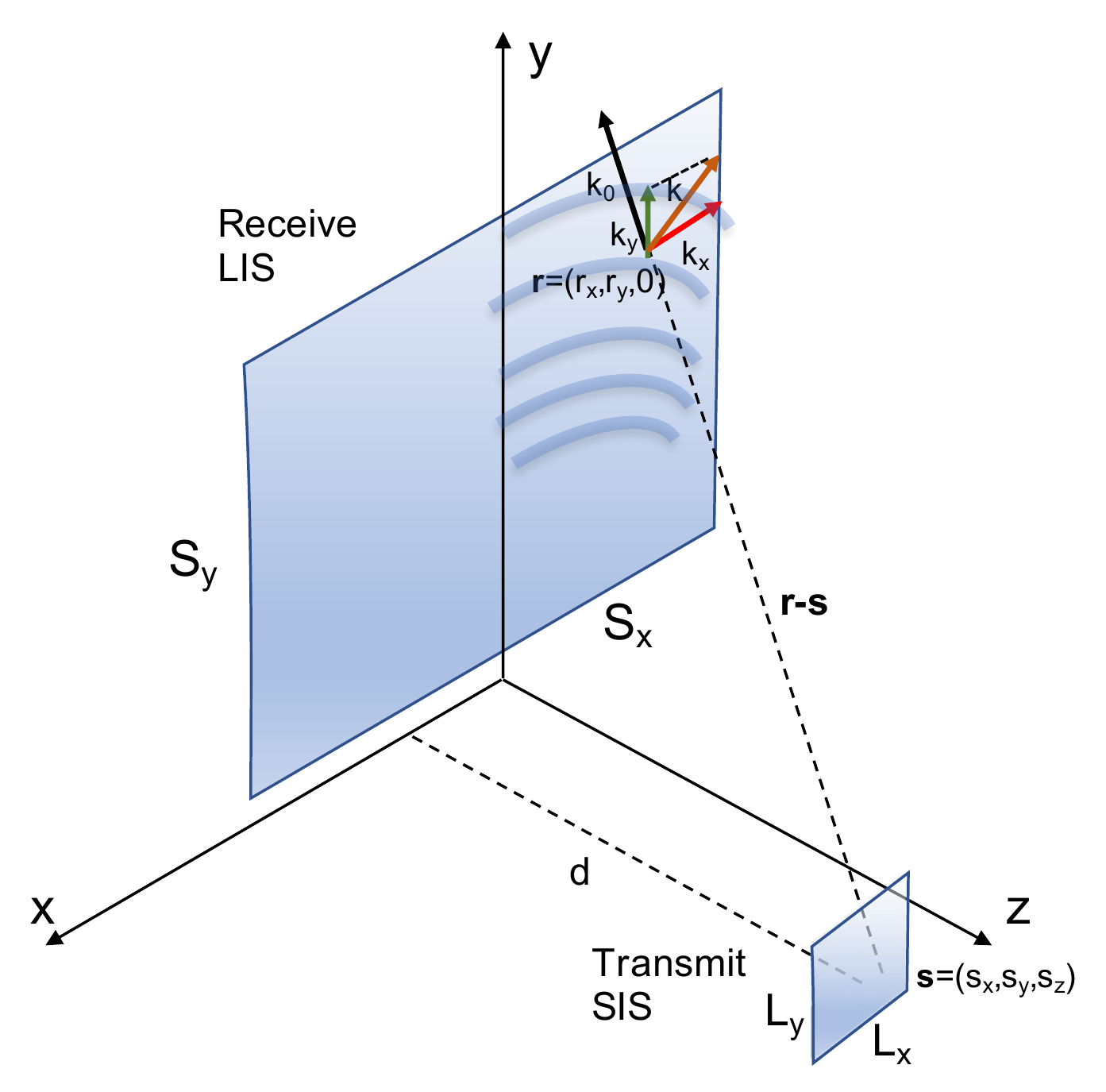}}
\caption{Geometric configuration between SIS and LIS.}
\label{Fig:Geometry}
\end{figure}

\section{Power Gain Between Large and Small-medium Intelligent Surfaces}
\label{Sec:Power}

The coupling intensity for any generic geometric configuration of antennas can be easily computed by solving \eqref{eq:coupling2} numerically.
Nevertheless, closed-form expressions can be obtained for some relevant cases from which  interesting considerations can be derived.

Consider a transmit  \ac{MIS}/{SIS} and a  receive \ac{LIS} at distance $d$. This situation  is expected to be common  in practice where the \ac{MIS}/{SIS} antenna might be embedded, for instance, into a smartphone or on top of a car, whereas the \ac{LIS} coats a wall of a building (as in  Fig. \ref{Fig:Scenario}). 
Without loss of generality, the receive \ac{LIS}  is deployed along the $xy$-plane at $z=0$, therefore the generic point on the surface is represented by the coordinates $\boldr=(\rx,\ry,0) \in \Sr$ (see Fig. \ref{Fig:Geometry}).
Denote with $\bolds=(\sx,\sy,\sz) \in \St$ the coordinates of  the generic point source of the transmit surface $\St$.
The centers and sizes of the transmit and receive intelligent surfaces  are, respectively, $\bolds_0=(\xo,\yo,d)$, $(\Lx,\Ly)$ and  $\boldr_0=(0,0,0)$, $(\Sx,\Sy)$. The corresponding areas are $\At=\Lx \, \Ly$ and $\Ar=\Sx \, \Sy$. Since the transmit antenna is a \ac{MIS}/{SIS}, it is reasonable to assume that $\Lx, \Ly \ll d$, $\Lx \ll \Sx$, and  $\Ly \ll \Sy$. Contrarily, $\Sx$ and $\Sy$ may be of the same order of magnitude as $d$.
 
To calculate the link power gain between the transmit \ac{SIS} and the receive \ac{LIS}, one has to consider only the component   
 of the power integrand in \eqref{eq:coupling2} perpendicular to the surface, i.e., 
\begin{align} \label{eq:gain2}
	g&=\frac{1}{\lambda^2} \int_{\Sr} \int_{\St} \frac{(r_y-s_y)^2+(r_z-s_z)^2}{|\boldr - \bolds|^4} \versorp \cdot \hatbn\,  d \boldr \, d \bolds \nonumber \\
	& \simeq \frac{\At}{\lambda^2} \int_{\Sr}  \frac{\left ((r_y-y_0)^2+d^2 \right )\, d}{|\boldr - \bolds_0|^{5}} \,  d \boldr \, ,
\end{align}
where $\versorp=(\boldr - \bolds)/|\boldr - \bolds|$  is the direction of propagation, and $\hatbn=\hatbz$.  In \eqref{eq:gain2}, we have made the approximations $s_z \simeq d$ and  $|\boldr - \bolds|^2 \simeq |\boldr - \bolds_0|^2$, since the transmit antenna is small compared to the distance $d$. As a consequence, the result does not depend on \ac{SIS}' orientation but only on its area $\At$.
Equation \eqref{eq:gain2} can be solved in closed-form but the final expression is quite articulated and it does not provide important insights.  
Therefore, for the sake of space, we report here the result valid for $\bolds_0=(0,0,d)$ from which some interesting conclusions can be drawn. 
In this case, \eqref{eq:gain2} becomes
\begin{align}
g & =  \frac{\At}{\lambda^2} \int_{-\Sx/2}^{\Sx/2} \int_{-\Sy/2}^{\Sy/2} 
\frac{ d \left(\ry^2+d^2\right)  }{ \left(\rx^2+\ry^2+d^2\right)^{5/2}} \, d\rx \, d\ry  \nonumber \\
 & =  \frac{4 d\, \At \Sx}{3 \lambda^2}  \int_{-\Sy/2}^{\Sy/2} 
 \frac{  6 d^2+\Sx^2+6 \ry^2}{ \left(d^2+\ry^2\right) \left(4 d^2+\Sx^2+4 \ry^2\right)^{3/2}} \, d\ry \, ,
\end{align}
which gives
\begin{align} \label{eq:gain3}
g&=\frac{8 \At \left( \frac{\Sx \Sy \, d}{\left(\Sx^2+4 d^2 \right) \sqrt{\Sx^2+\Sy^2+4 d^2}}+\tan^{-1} \left(\frac{\Sx \Sy}{2 d \sqrt{\Sx^2+\Sy^2+4 d^2}}\right)   \right)  }{3\lambda^2}  .
\end{align}
For a square  \ac{LIS}, the last expression simplifies as follows
\begin{align} \label{eq:gainsquared}
	g&= \frac{4 \At}{3 \lambda^2} \left ( \frac{\sqrt{2F}}{ \sqrt{1 + 2 F}  (1 + 4 F)} + 2 \text{acot} \left (\sqrt{8F (1 + 2 F)} \right ) \right )  \, ,
\end{align}
where $F=d^2/\Ar$. It can be observed from \eqref{eq:gainsquared} that the gain is a function of relative geometric quantities, i.e., the normalized (to the wavelength)  transmit \ac{SIS}' area $\At$ and the ratio $F$. 

It is interesting to analyze the behavior of \eqref{eq:gainsquared} when the \ac{LIS} is extremely large compared to the distance $d$ ($F \rightarrow 0$), that is
\begin{align} \label{eq:Gainlimit}
  g^{(\text{large LIS})} = \frac{4 \pi \At }{3 \lambda^2}  \, ,
\end{align}
which becomes independent of the distance. Instead,   for large distances ($F \rightarrow \infty$), corresponding to the Fraunhofer far-field region, \eqref{eq:gainsquared} gives
\begin{align} \label{eq:Friis}
  g^{(\text{large }d)} = \frac{\At \Ar  }{ \lambda^2 d^2}  \, .
\end{align}
The latter is the result found by Miller \cite{Miller:00} (reported in \eqref{eq:gainMiller} with a different normalization factor) when considering thin volumes and it is nothing else than the well-known Friis' formula. 
In fact, if one defines $\Gi=\lambda^2/(4 \pi d)^2$, $\Gt= 4 \pi \At/\lambda^2$, and $\Gr=4 \pi \Ar  / \lambda^2$,  respectively, the isotropic free-space channel gain, the gain of the transmit  and receive antennas considered as aperture antennas,  it is $g^{(\text{large }d)} =\Gt \, \Gr \, \Gi$ \cite{BalB:16}.

It is worth to notice that the comparison between  \eqref{eq:Gainlimit} and \eqref{eq:Friis} puts in evidence the limitation of classical path-loss formulas when using \acp{LIS}. In fact, from \eqref{eq:Friis} one could draw the conclusion that by increasing the size of both the transmit and receive antenna it is possible to increase the link gain to any desired level. Instead,  \eqref{eq:Gainlimit}  tells that this is possible only up to a certain extent, i.e., until the size of one of the two antennas becomes very large so that the system works in the near-field region. In that region, the link gain is limited by the (normalized) area of the smallest of the two antennas.  This result is a direct consequence of the diffraction effect of  \ac{e.m.} waves. 

Equation \eqref{eq:gain2} and, in particular, \eqref{eq:gain3} and \eqref{eq:gainsquared} represent simple design formulas useful to characterize the link budget in \ac{LIS}-based communications without resorting to \ac{e.m.} extensive simulations.

\section{Communication \ac{DoF} between Intelligent Surfaces }
\label{Sec:DoF}

In this section we derive approximate expressions for the communication \ac{DoF} between a transmit \ac{SIS} and a receive \ac{LIS} antenna   following 2D sampling theory arguments. The accuracy of such expressions, with respect to the actual \ac{DoF} value from the eigenfunction problems in Sec. \ref{Sec:Problem}, is addressed in the numerical results. 

With reference to Fig. \ref{Fig:Geometry}, the wave originated by the point source $\bolds$ has wavenumber $k_0$ in the radial direction $\boldr-\bolds$ between the point source and the generic point $\boldr$ on the receive (observation) surface $\Sr$.
Contrarily to what happens in the 1D coordinate system, where a linear transformation never changes or generates new frequency components, when moving to 2D and 3D coordinate systems, it may happen that the observed wavenumber is different from $k_0$   if the observation direction is different from that of $\boldr-\bolds$.
More specifically, along the $x$ and $y$ directions of the receive surface, the observed wave is characterized by wavenumber  
\begin{align}
 	\boldk(\boldr,\bolds) & =k_0 \left ( \versorp-\versorn \, (\versorp \cdot \versorn)  \right )=(k_x(\boldr,\bolds),k_y(\boldr,\bolds))\, ,
\end{align}
where $\versorp=(\boldr - \bolds)/|\boldr - \bolds|$,  and $\versorn$ is the unit vector perpendicular to the surface in the point $\boldr$, so that
\begin{align}
 	k_x(\boldr,\bolds)& =k_0 \frac{r_x-s_x}{\sqrt{(r_x-s_x)^2+(r_y-s_y)^2+s_z^2}} \nonumber \\
	k_y(\boldr,\bolds) &=k_0 \frac{r_y-s_y}{\sqrt{(r_x-s_x)^2+(r_y-s_y)^2+s_z^2}} \, .
\end{align}

Consider now an infinitesimal surface $d\boldr$ centered in $\boldr$. The received wave observed in $d\boldr$ can be seen as a two-dimensional signal whose local bandwidth changes slowly with $\boldr - \bolds$, and it is approximatively constant in $d \boldr$.  
The local bandwidth in the wavenumber domain observed in $d\boldr$ is the maximum wavenumber spread related to all point sources in $\St$. Specifically it is
\begin{align} \label{eq:B}
B(\boldr)=& \frac{1}{4} \text{area}   \left [ \boldk(\boldr,\bolds)   \right ]_{\bolds \in \St}\, ,
\end{align}
where the operator  $\text{area} [ \cdot ]_{\bolds \in \St}$ returns the area of the region in the complex plane spanned by the function $\boldk(\boldr,\bolds)$ when parameter $\bolds$ varies in $\St$. 

Considering that the  number of requested samples at Nyquist rate (i.e., the \ac{DoF}) to represent  a 2D signal of spatial bandwidth $B$ in an area $S$ is  equal to\footnote{ Since a signal cannot be limited in both domains, this expression represents an approximation. An extensive discussion on this subject  can be found in \cite{Slepian:83}.} $B\, S/\pi^2$,  the \ac{DoF} of the signal ``projected" onto $\Sr$ results
\begin{align} \label{eq:D}
D =& \frac{1}{\pi^2} \int_{\Sr} B(\boldr) \, d\boldr    \, .
\end{align}

In the next section we will make \eqref{eq:D} particular to some \ac{LIS} configurations with the purpose to derive simple expressions of the  \ac{DoF}  and obtain some interesting insights.

\subsection{DoF of Communicating Parallel LIS and SIS}
\label{sec:parallel}

For parallel intelligent surfaces, a way to compute \eqref{eq:B} is to approximate the curve delimiting the area $\left [ \boldk(\boldr,\bolds)   \right ]_{\bolds \in \St}$   with a quadrilateral having vertices given by   $k_x^{(i)}(\boldr)=k_x\left (\boldr,\bolds^{(i)} \right )$, $k_y^{(i)}(\boldr)=k_y\left (\boldr,\bolds^{(i)} \right )$, $i=1,2, \ldots , 5$, with $\bolds^{(1)}=(\xo-\Lx/2,\yo-\Ly/2,d)$, $\bolds^{(2)}=(\xo+\Lx/2,\yo-\Ly/2,d)$, $\bolds^{(3)}=(\xo-\Lx/2,\yo+\Ly/2,d)$, $\bolds^{(4)}=(\xo+\Lx/2,\yo+\Ly/2,d)$, $\bolds^{(5)}=\bolds^{(1)}$, and then by applying the Gauss' formula 
\begin{align} \label{eq:Aa}
	A(\boldr) \simeq \frac{1}{2} \left | \sum_{i=1}^4 \left ( k_x^{(i)}(\boldr) \, k_y^{(i+1)}(\boldr) - k_x^{(i+1)}(\boldr) \, k_y^{(i)}(\boldr) \right  ) \right | \, . 
\end{align}

From \eqref{eq:B}, \eqref{eq:D} and \eqref{eq:Aa} it follows that
 \begin{align}  \label{eq:DParallel}
 D^{||} \simeq  & \frac{1}{4 \pi^2} \int_{\Sr} A(\boldr) \, d \boldr \, .
 \end{align} 
 
Unfortunately, \eqref{eq:DParallel} does not admit a closed-form expression in general. However, even though it requires the evaluation of a two-folded integral, its numerical computation is very fast and it does not pose any particular issue compared to the numerical complexity of the eigenfunction problems \eqref{eq:xi1} and \eqref{eq:xi2}.

Nevertheless, it could be of interest to derive closed-form expressions of \eqref{eq:DParallel} for some significant cases.
Specifically, since $\Lx, \Ly \ll d$, setting $\xo=\yo=0$,  \eqref{eq:DParallel} gives (details are reported in Appendix B)
\begin{align} \label{eq:DParallel1}
D^{||} \simeq &\frac{2 \Lx \Ly}{\lambda^2}  \left(\frac{\Sx \tan^{-1}\left(\frac{\Sy}{\sqrt{4 d^2+\Sx^2}}\right)}{\sqrt{4 d^2+\Sx^2}}+\frac{\Sy \tan^{-1}\left(\frac{\Sx}{\sqrt{4 d^2+\Sy^2}}\right)}{\sqrt{4 d^2+\Sy^2}}\right) \, .
\end{align}

For  $d\gg \Sx,\Sy$, i.e., in the far-field region,  it is 
\begin{align} \label{eq:Dlarged}
	D_{\text{large}}^{||}=\frac{ \At \Ar}{\lambda^2 d^2} \, ,
\end{align}
which gives  \eqref{eq:Dmiller} derived in \cite{Miller:00}.

The limit of \eqref{eq:DParallel1} for $\Sx,\Sy \rightarrow \infty$, i.e., very large surfaces, is 
\begin{align} \label{eq:Dlim}
	D_{\text{asympt}}^{||}=\frac{\pi \, \Lx \, \Ly}{\lambda^2}=\frac{\pi \At}{\lambda^2} \, .
\end{align}
Equation \eqref{eq:Dlim} indicates that the maximum \ac{DoF} depends only on the area of the transmit surface (normalized to the square half-wavelength), i.e., the area of the smallest of the 2 antennas, and  it represents the ultimate \ac{DoF} limit which is  independent of the distance.   
This result is reminiscent of the \ac{DoF} in \ac{MIMO} systems when the channel matrix is full rank, i.e., in the presence of rich multipath \cite{TseVis:B05}.
Unfortunately, in \ac{LOS} channel condition, the rank of the \ac{MIMO} channel matrix is 1, and hence $D=1$ (only beamforming gain is present).
Instead, result \eqref{eq:DParallel1} indicates that with a \ac{LIS} one can obtain \ac{DoF}  larger than 1 even in \ac{LOS}.
Having large \ac{DoF} in  \ac{LOS} could  significantly increase the link capacity according to \eqref{eq:capacitygain}, especially at millimeter waves or in the THz band where the multipath is not rich or could be dominated by the \ac{LOS} component.

\subsection{DoF of Communicating Perpendicular \ac{LIS} and \ac{SIS}}

Consider now a transmit surface along the plane $xz$ with coordinates  $\bolds=(\sx,y_0,\sz)\in \St$ and a perpendicular receive \ac{LIS} at distance 
 $d$ with coordinates  $\boldr=(\rx,\ry,0)\in \Sr$. The centers and sizes of the transmit and receive intelligent surfaces  are, respectively, $\bolds_0=(\xo,\yo,d)$, $(\Lx,\Lz)$ and $(0,0,0)$, $(\Sx,\Sy)$. The corresponding areas are $\At=\Lx \, \Lz$ and $\Ar=\Sx \, \Sy$.

Following a similar approach as in Sec. \ref{sec:parallel},  by setting $\bolds^{(1)}=(\xo-\Lx/2,\yo,d-\Lz/2)$, $\bolds^{(2)}=(\xo+\Lx/2,\yo,d-\Lz/2)$, $\bolds^{(3)}=(\xo-\Lx/2,\yo,d+\Lz/2)$, $\bolds^{(4)}=(\xo+\Lx/2,\yo,d+\Lz/2)$, it is
\begin{align} \label{eq:DPerpendicular}
D^{\bot} \simeq \frac{2 \Lx \Lz  \left(\sqrt{4 d^2+\Sy^2} \cot ^{-1}\left(\frac{2 d}{\Sx}\right)-2 d \tan ^{-1}\left(\frac{\Sx}{\sqrt{4 d^2+\Sy^2}}\right)\right)
}{\lambda^2 \sqrt{4 d^2+\Sy^2}} \,  . 
\end{align}

For  $d\gg \Sx,\Sy$ one gets
\begin{align} \label{eq:DPerpendicularLarge}
D_{\text{large}}^{\bot} = \frac{ \At \Ar \Sy}{4 \lambda^2 d^3} \, , 
\end{align}
which, compared to  \eqref{eq:Dlarged} valid for parallel surfaces, denotes a dependence on the ratio $\Sy/d$. Such a term contributes to increase the \ac{DoF} when the \ac{LIS} is tall and hence is capable to ``see" better the transmit surface lying on the horizontal plane.  

The limit of \eqref{eq:DPerpendicular} for $\Sx,\Sy \rightarrow \infty$, i.e., very large surface, is 
\begin{align} \label{eq:DlimP}
	D_{\text{asympt}}^{\bot}=\frac{ \pi \At}{\lambda^2} \, ,  
\end{align}
that is, the same as parallel surfaces.

\begin{figure}[t]
\centerline{\includegraphics[width=1\columnwidth]{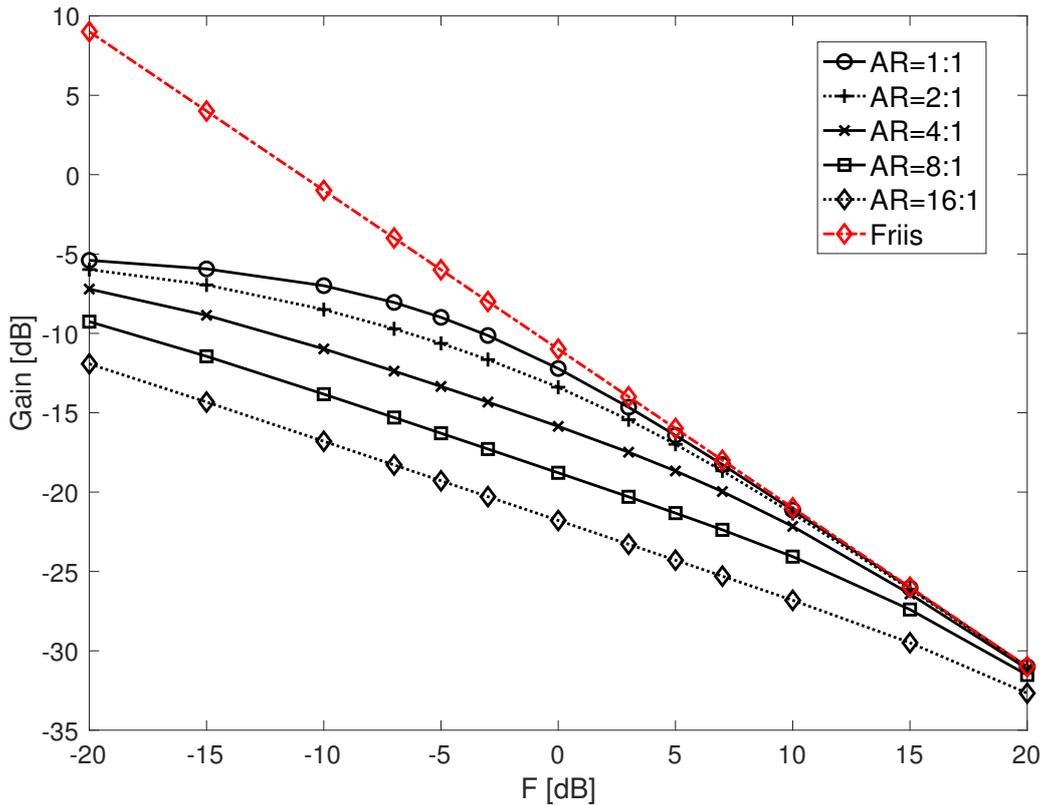}}
\caption{Normalized gain vs $F=d^2/\Ar$ of a SIS-LIS link. }
\label{Fig:Gain}
\end{figure}

\section{Numerical results}
\label{Sec:NumericalResults}

In this section, we  present some numerical examples with the purpose to illustrate the potential advantages in communicating with \acp{LIS}  and to assess the validity of the method proposed to compute the \ac{DoF}.

In Fig. \ref{Fig:Gain}, the link gain between a \ac{SIS} communicating with a \ac{LIS} using  \eqref{eq:gain3}, normalized to   $\Gt=\At \, 4\pi/\lambda^2$, is shown as a function of $F$ and for different values of \ac{LIS}' aspect ratio $\AR=\Sx:\Sy$.
Notice that this plot does not depend on $\lambda$, on the absolute distance between the intelligent surfaces, and the dimension of the receive \ac{LIS}, but only on the relative quantities $F=d^2/\Ar$ and $\AR$.
When the size of the \ac{LIS} is comparable or larger than the distance from the transmitter (small $F$), near-field effects become dominant leading to a saturation of the link gain toward the limit value   \eqref{eq:Gainlimit}. This can be ascribed to diffraction effects, which make the commonly used antenna aperture formula, according to which the antenna gain is proportional to the geometric area, no longer valid. From Fig. \ref{Fig:Gain}, it can  also be noticed that the best geometric shape is the square one ($\AR=1:1$). For comparison, the gain obtained using the Friis' formula \eqref{eq:Friis} is also shown, from which it is evident that it fails in modeling the link budget when \acp{LIS} are used, especially for low $F$. 

\begin{figure}[t]
\centerline{\includegraphics[width=1\columnwidth]{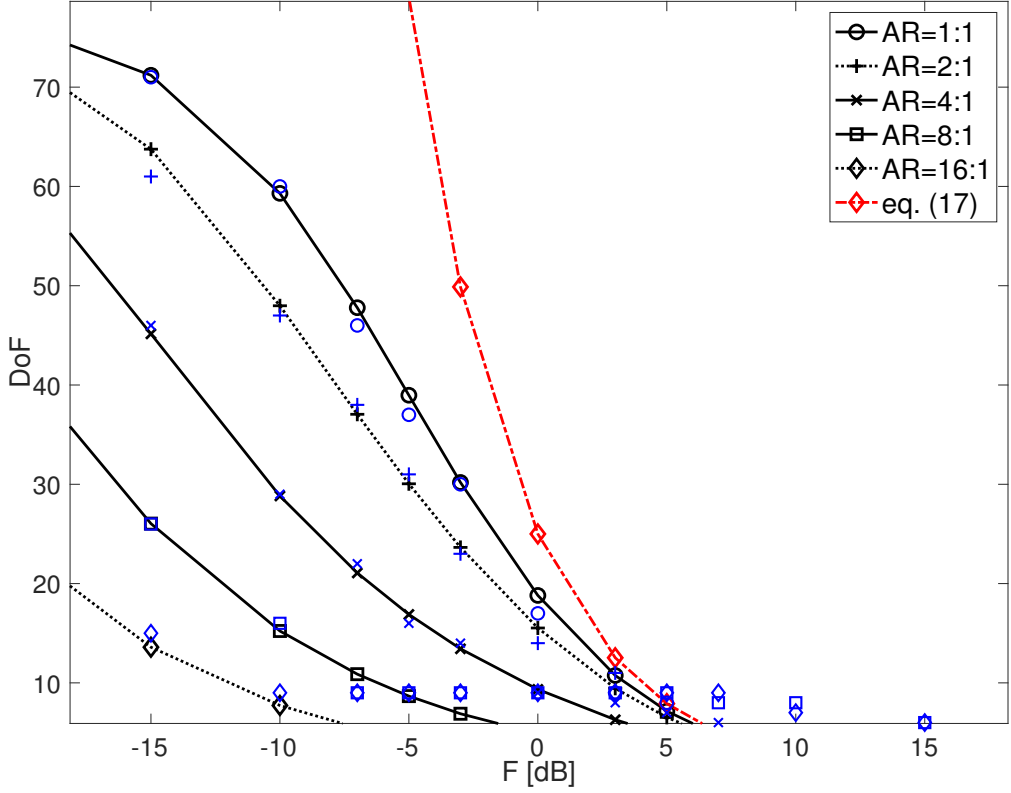}}
\caption{\ac{DoF} vs $F=d^2/\Ar$ for parallel surfaces. $\Ar=25\,$cm$^2$, $f_c=28\,$GHz. Blue markers refer to numerical solution to the eigenfunction problems.}
\label{Fig:DoF}
\end{figure}

Now we investigate the \ac{DoF} available when a \ac{LIS} and a \ac{SIS} are communicating in the near- and far-field. 
Fig. \ref{Fig:DoF} shows the \ac{DoF} in \eqref{eq:DParallel1} related to parallel surfaces as a function of $F$ for different values of $\AR$, with  $\lambda=1\,$cm ($f_c=28\,$GHz), and $5\times5\,$cm$^2$ \ac{LIS}  ($\At=25\,$cm$^2$).\footnote{Although the values obtained from  \eqref{eq:DParallel1} should be rounded to the nearest integer value larger or equal to 1, here the continuous version is plotted to easy the reading.}

For low $F$ (very large \ac{LIS}), the \ac{DoF} saturates to the limit value given by \eqref{eq:Dlim},  in this case equal to $78$.
As far as the Fraunhofer far-field regime is approached (large $F$), the \ac{DoF} tends to one, as in conventional \ac{MIMO} systems in \ac{LOS} condition where only the beamforming gain is present. Again, the best \ac{LIS}  configuration is given by the square shape ($\AR=1:1$).
The result obtained using \eqref{eq:Dmiller} by \cite{Miller:00} is also reported. It is evident how this expression, valid for antennas at distances much larger than their dimension, is not accurate for small $F$ and it is not able to capture the effect of the aspect ratio of the \ac{LIS}.

 \begin{figure}[t]
\centering
\subfigure[$n=1$]{\includegraphics[clip,width=0.49\columnwidth]{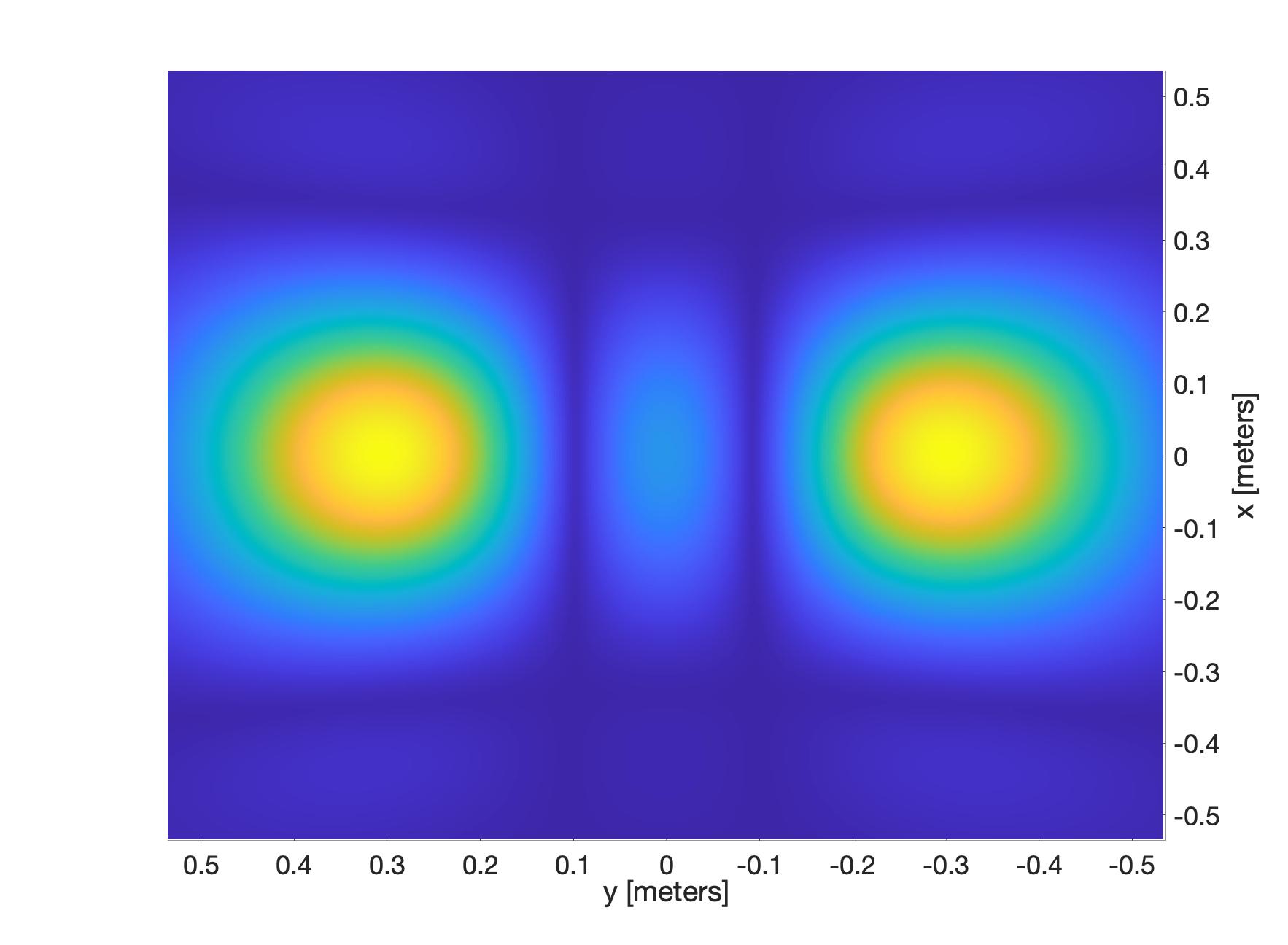}}
\subfigure[$n=2$]{\includegraphics[clip,width=0.49\columnwidth]{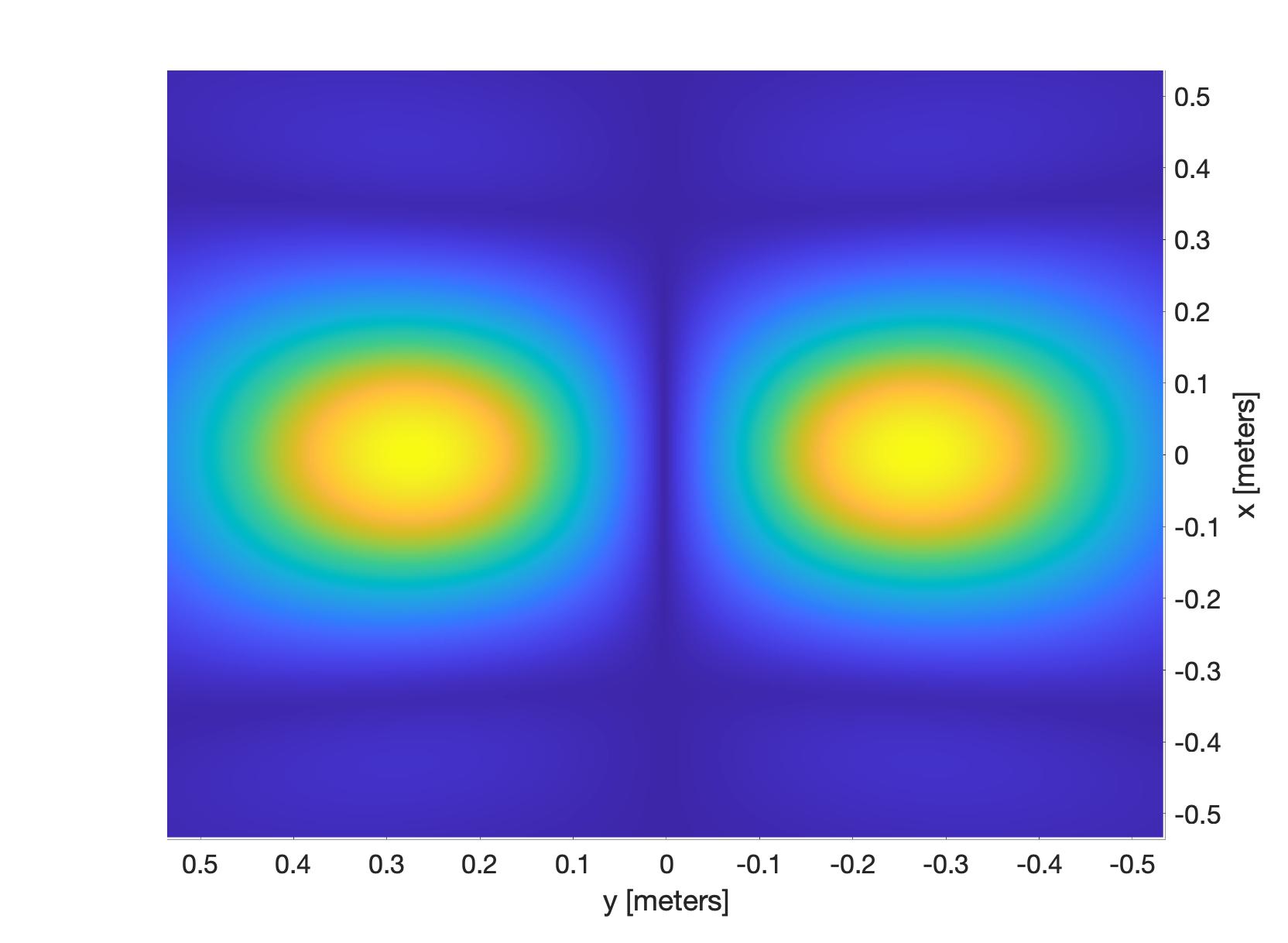}}
\subfigure[$n=3$]{\includegraphics[clip,width=0.49\columnwidth]{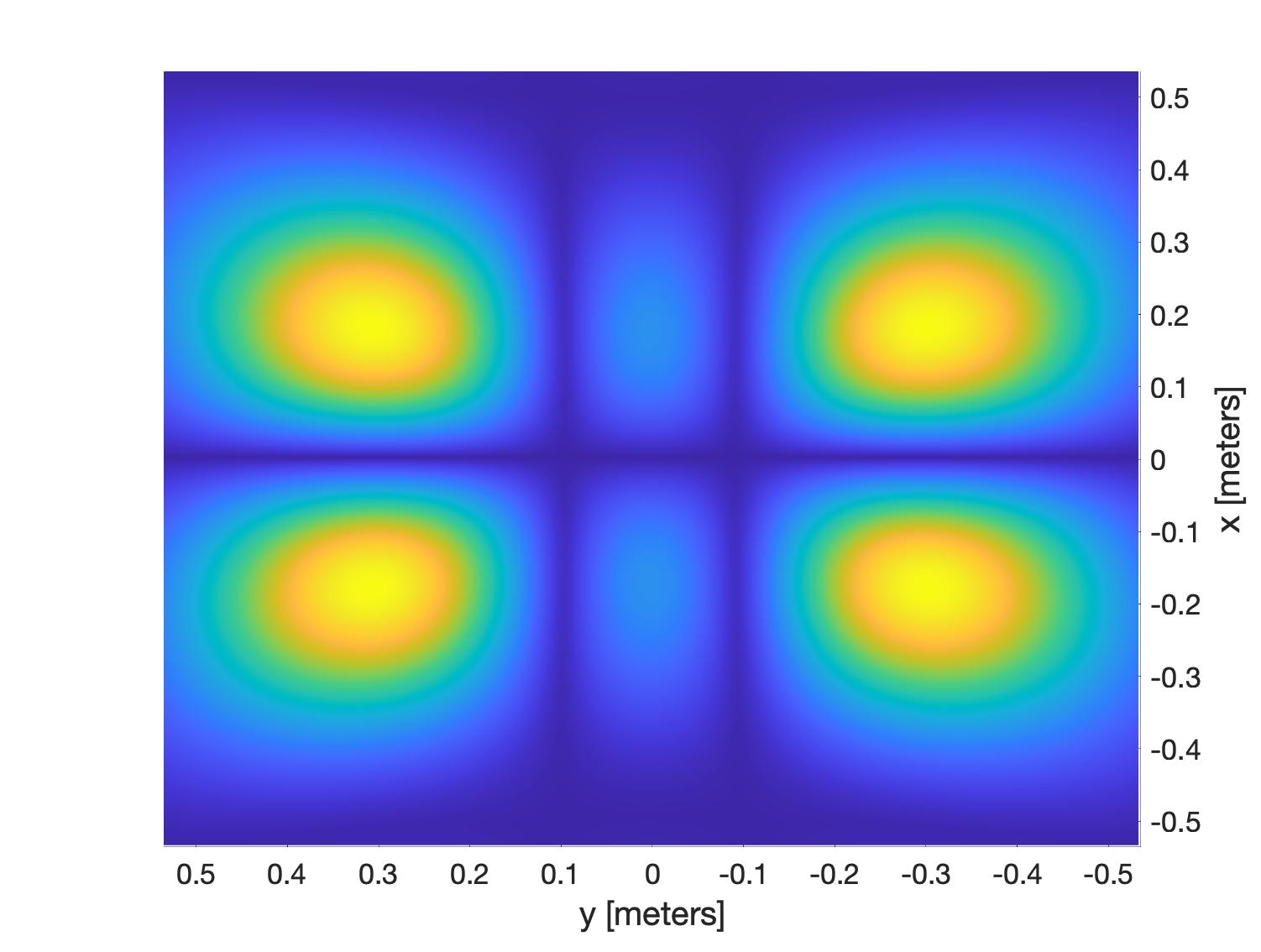}}
\subfigure[$n=4$]{\includegraphics[clip,width=0.49\columnwidth]{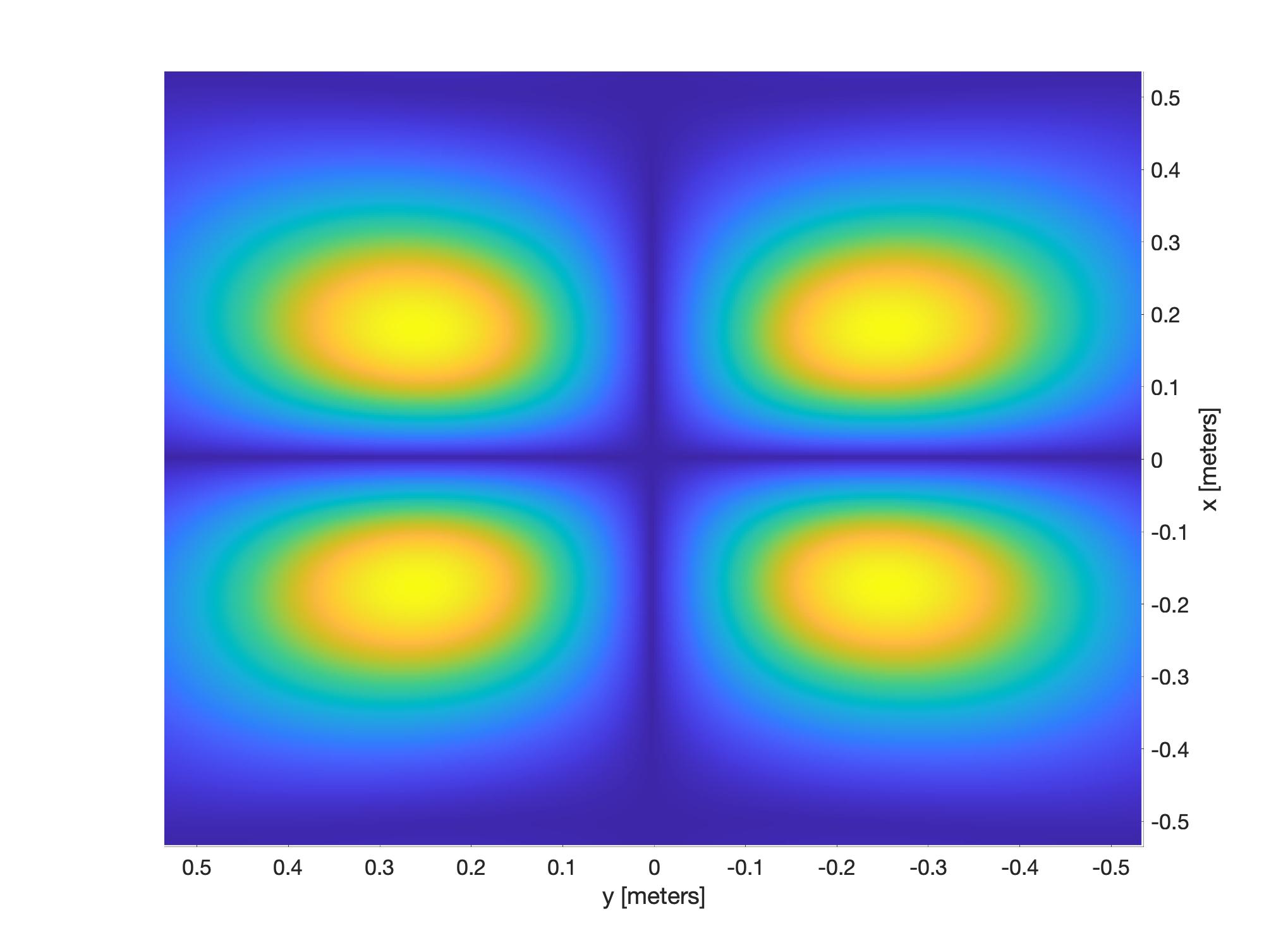}}
\caption{Amplitude of the $x$-component of eigenfunctions $\{\bpsi_n(\boldr)\}$ (receive LIS).}
\label{Fig:eigenfunctions_mag}
\end{figure}

\begin{figure}[t]
\centering
\subfigure[$n=1$]{\includegraphics[clip,width=0.49\columnwidth]{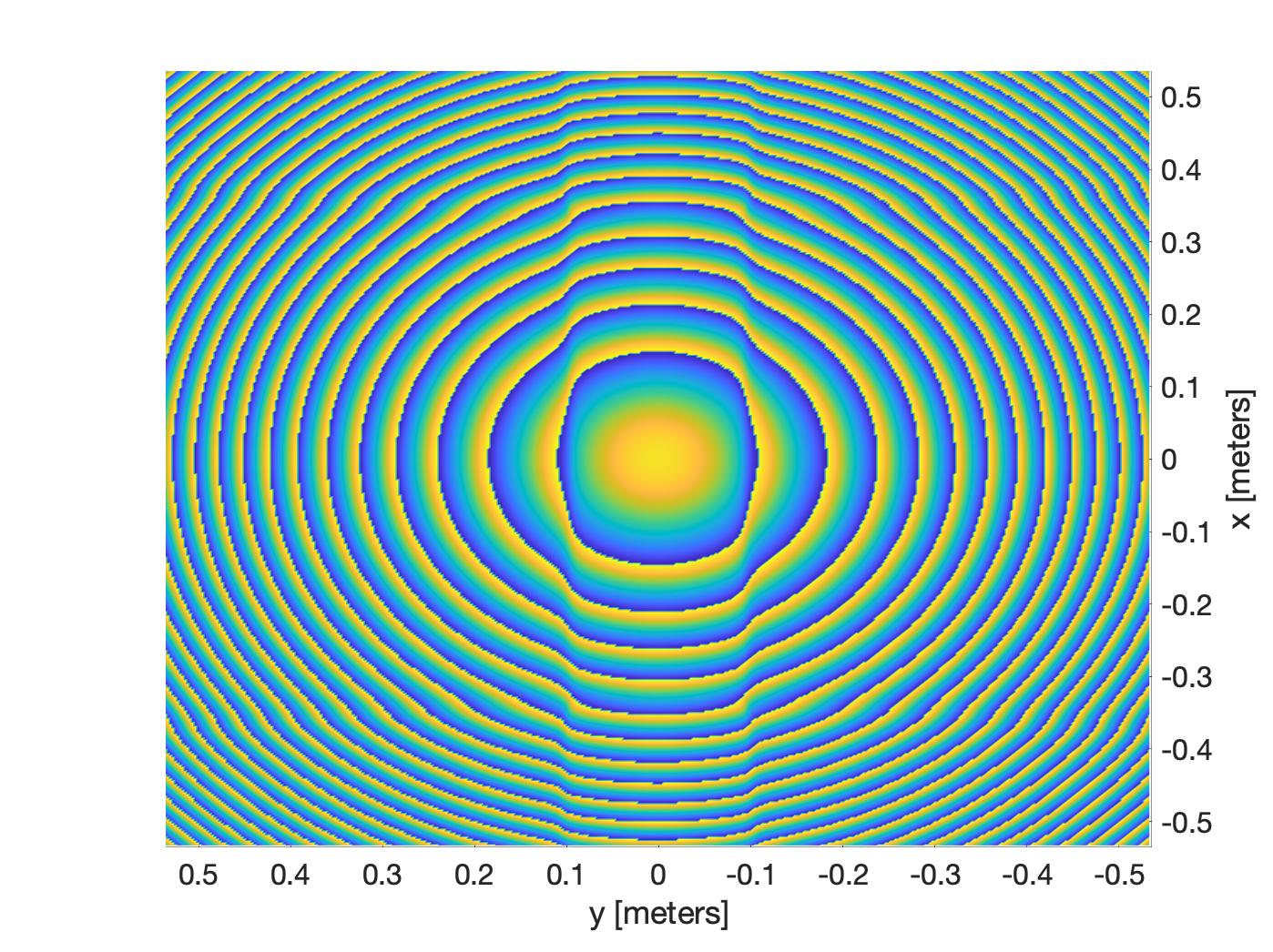}}
\subfigure[$n=2$]{\includegraphics[clip,width=0.49\columnwidth]{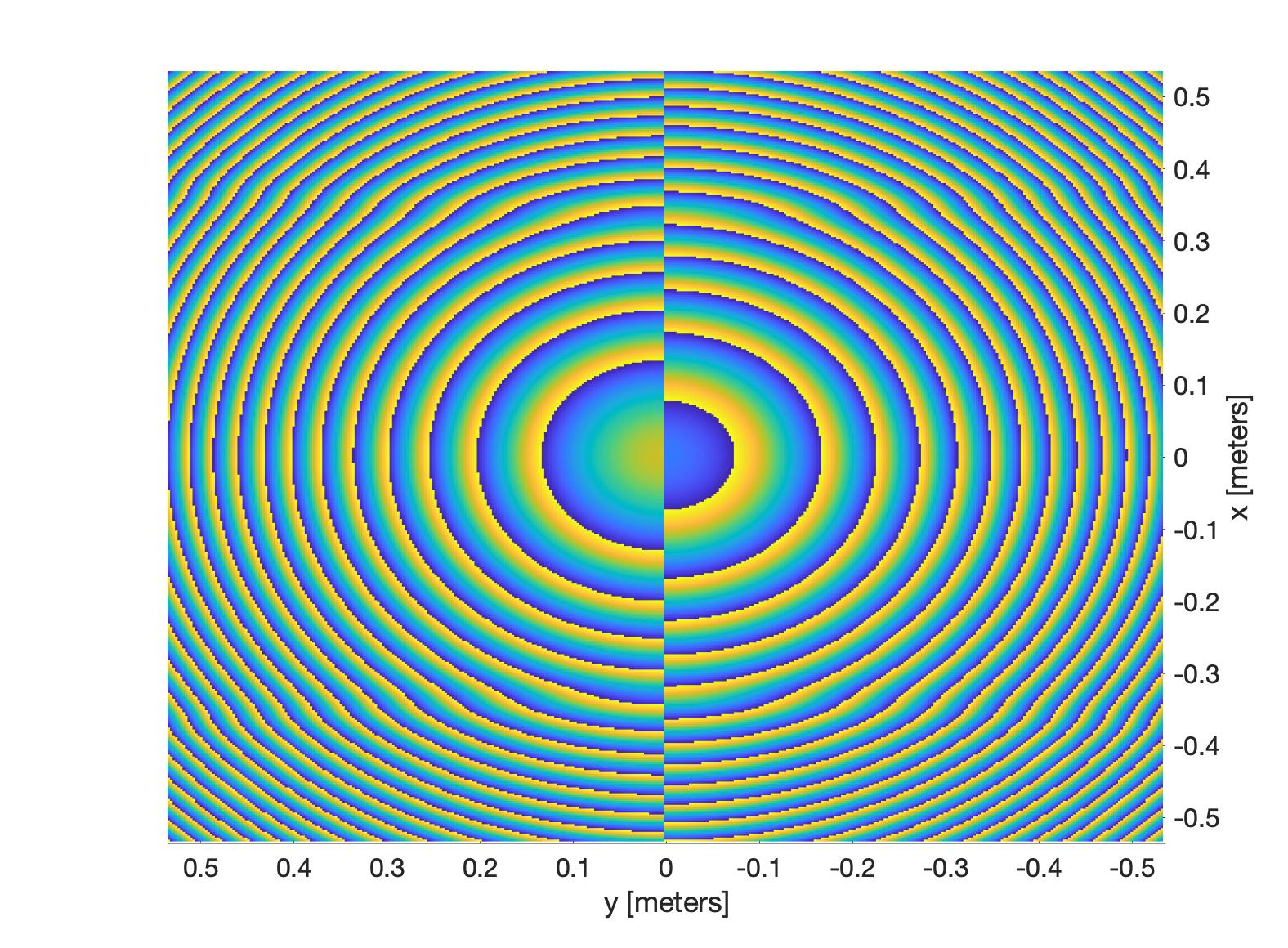}}
\subfigure[$n=3$]{\includegraphics[clip,width=0.49\columnwidth]{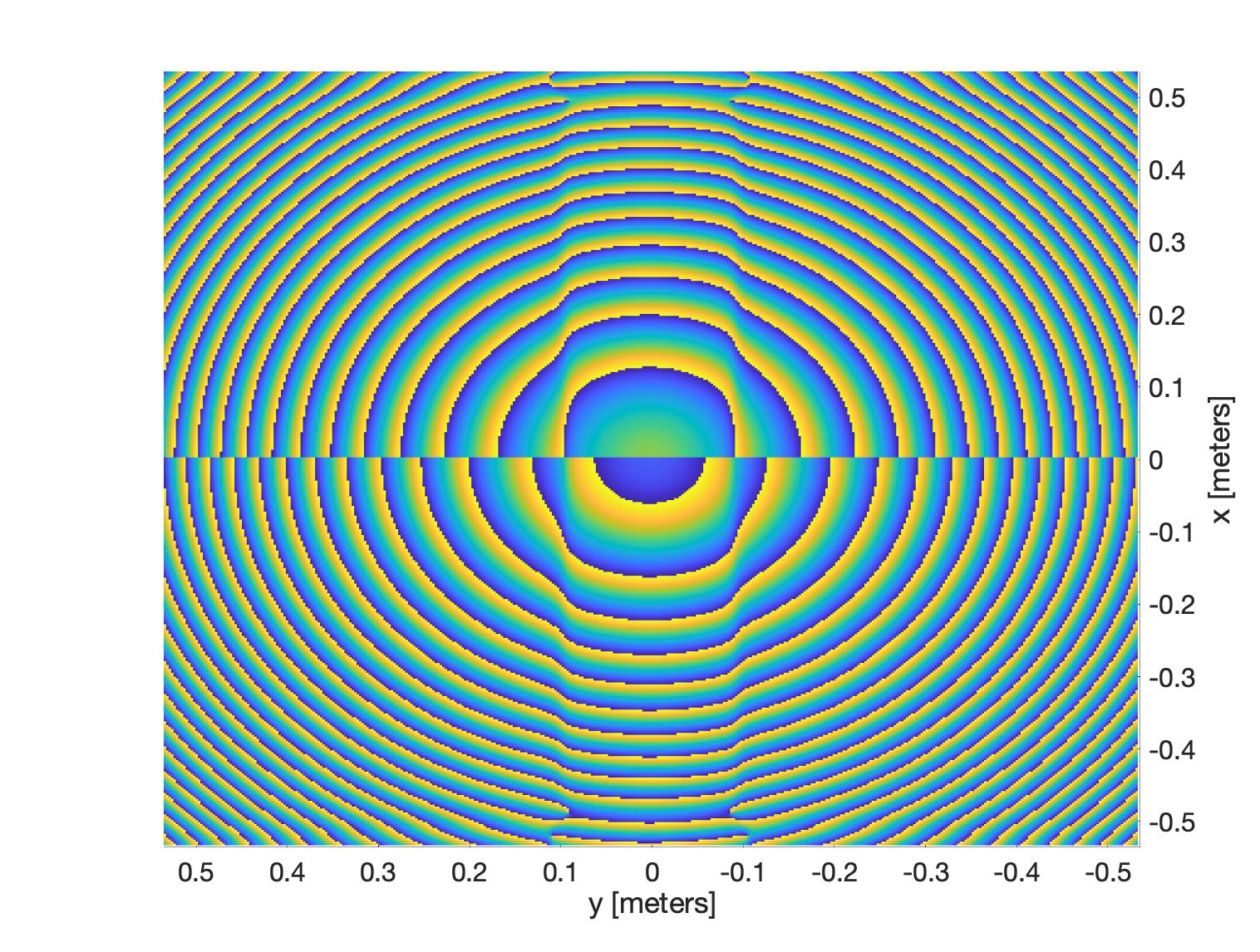}}
\subfigure[$n=4$]{\includegraphics[clip,width=0.49\columnwidth]{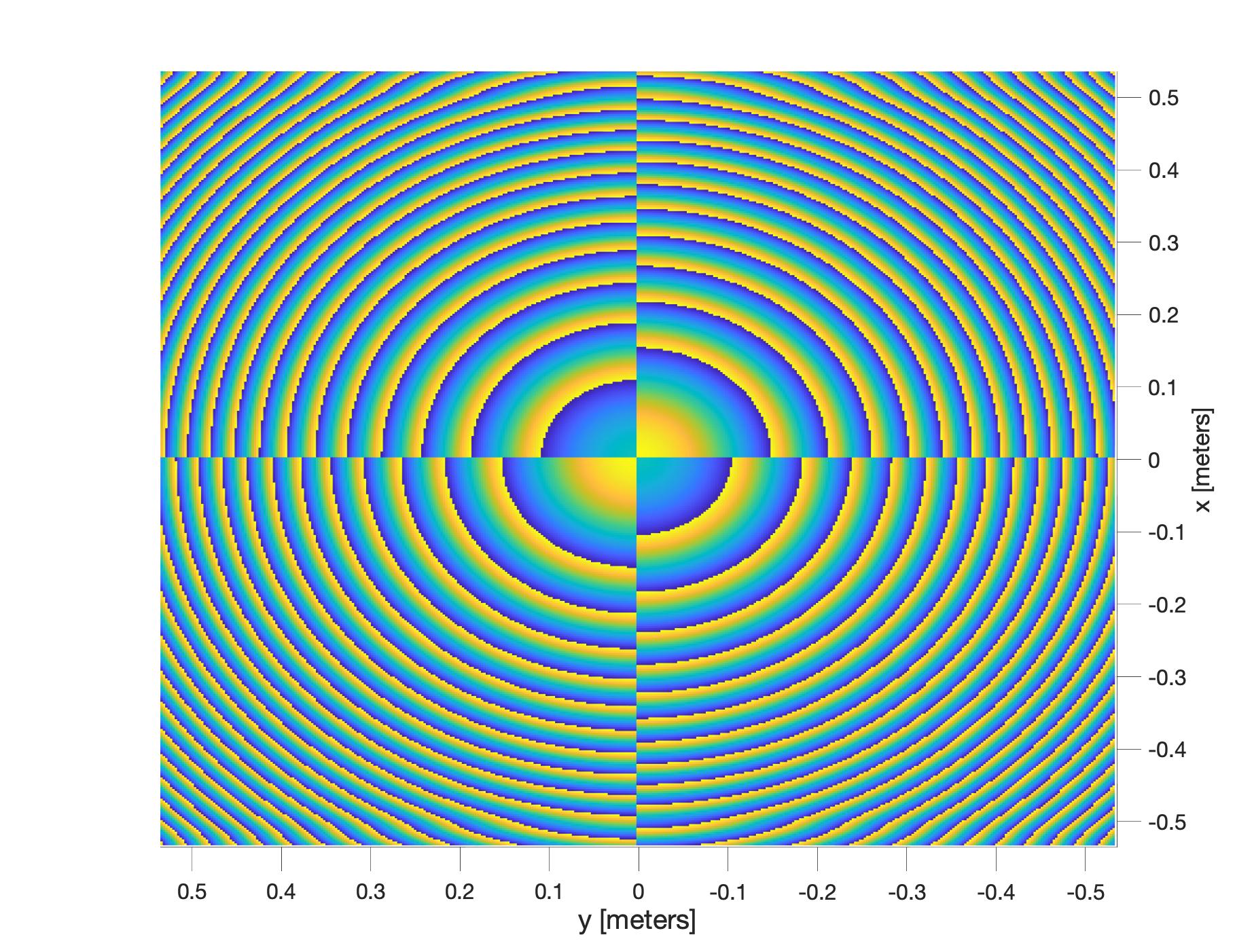}}
\caption{Phase of the $x$-component of eigenfunctions $\{\bpsi_n(\boldr)\}$ (receive LIS).}
\label{Fig:eigenfunctions_phase}
\end{figure}

In order to validate the approach proposed in Sec. \ref{Sec:DoF}, results have been compared to those obtained by solving numerically the eigenfunction problems in Sec. \ref{Sec:Problem}.
 To this purpose,  different numerical approximation methods exist (e.g., Galerkin's method) \cite{PieMil:00}. Among them, we considered the following one: we decomposed each surface in very small square patches of side $\Delta=\lambda/16$ and we considered them as piece-wise constant basis functions for the surfaces. In this way,   the eigenfunction problems can be approximated into  a singular-value decomposition problem with dimension $\Ar/\Delta^2 \times \At/\Delta^2$. Unfortunately, such a method becomes intractable as soon as the surfaces become large compared to $\lambda$ due to the corresponding huge dimension of the matrix to decompose. 
To make the computation time affordable, we considered a \ac{MIS} with $\Ar=1\,$m$^2$.  
The  \ac{DoF} has been computed by considering the largest eigenvalues within a tolerance of $3\,$dB.
  Results are plotted in Fig. \ref{Fig:DoF} (blue markers) and show a good agreement with the model developed in Sec. \ref{Sec:DoF}, especially for small $F$.  
  For large $F$, there are some discrepancies, but the fact that our results are consistent with the analytical expression \eqref{eq:Dmiller}, which is accurate for large $F$, generates the suspect of numerical evaluation issues caused by the singular-value decomposition of huge likely ill-posed matrices.

\begin{figure}[t]
\centering
\subfigure[$n=1$]{\includegraphics[clip,width=0.49\columnwidth]{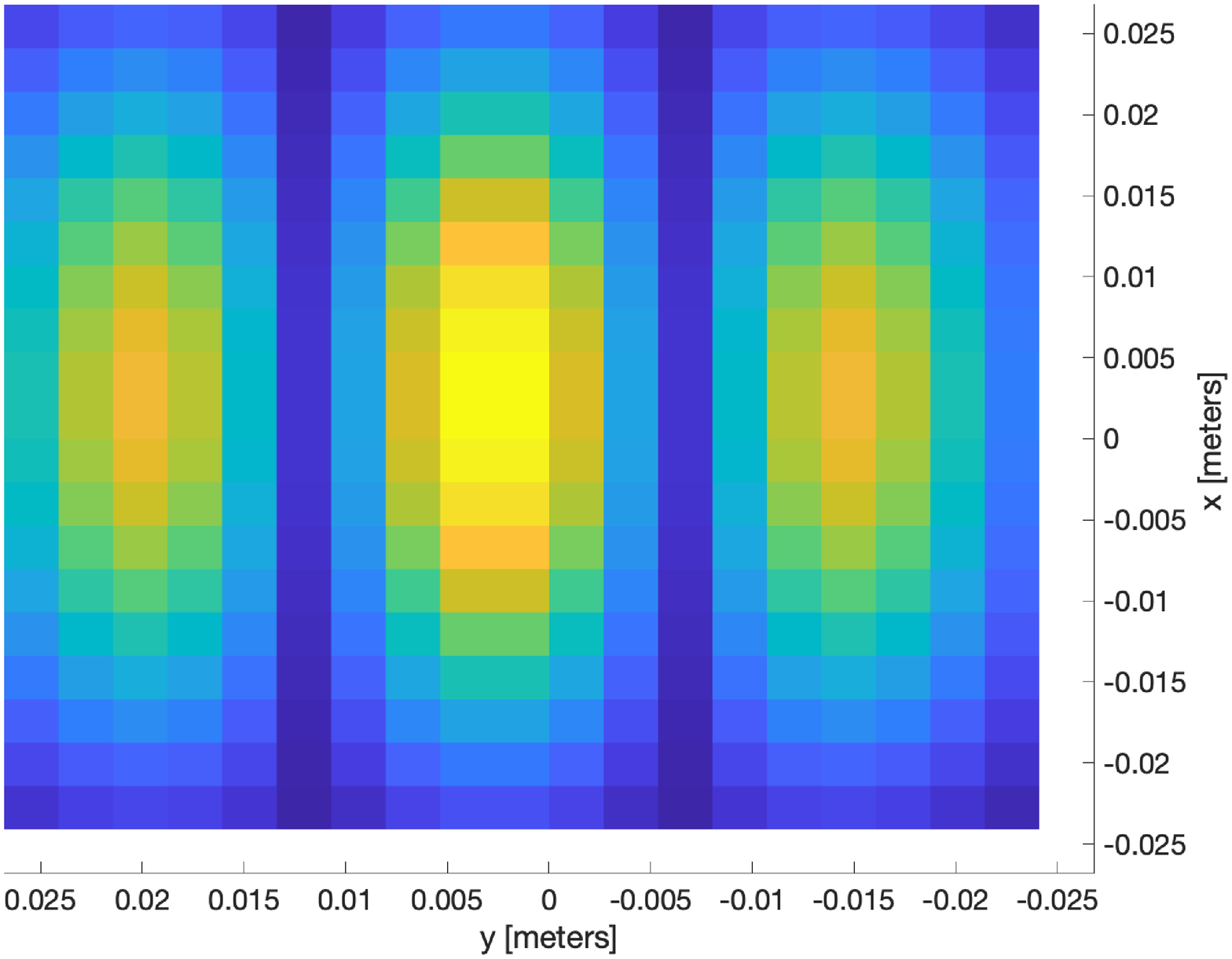}}
\subfigure[$n=2$]{\includegraphics[clip,width=0.49\columnwidth]{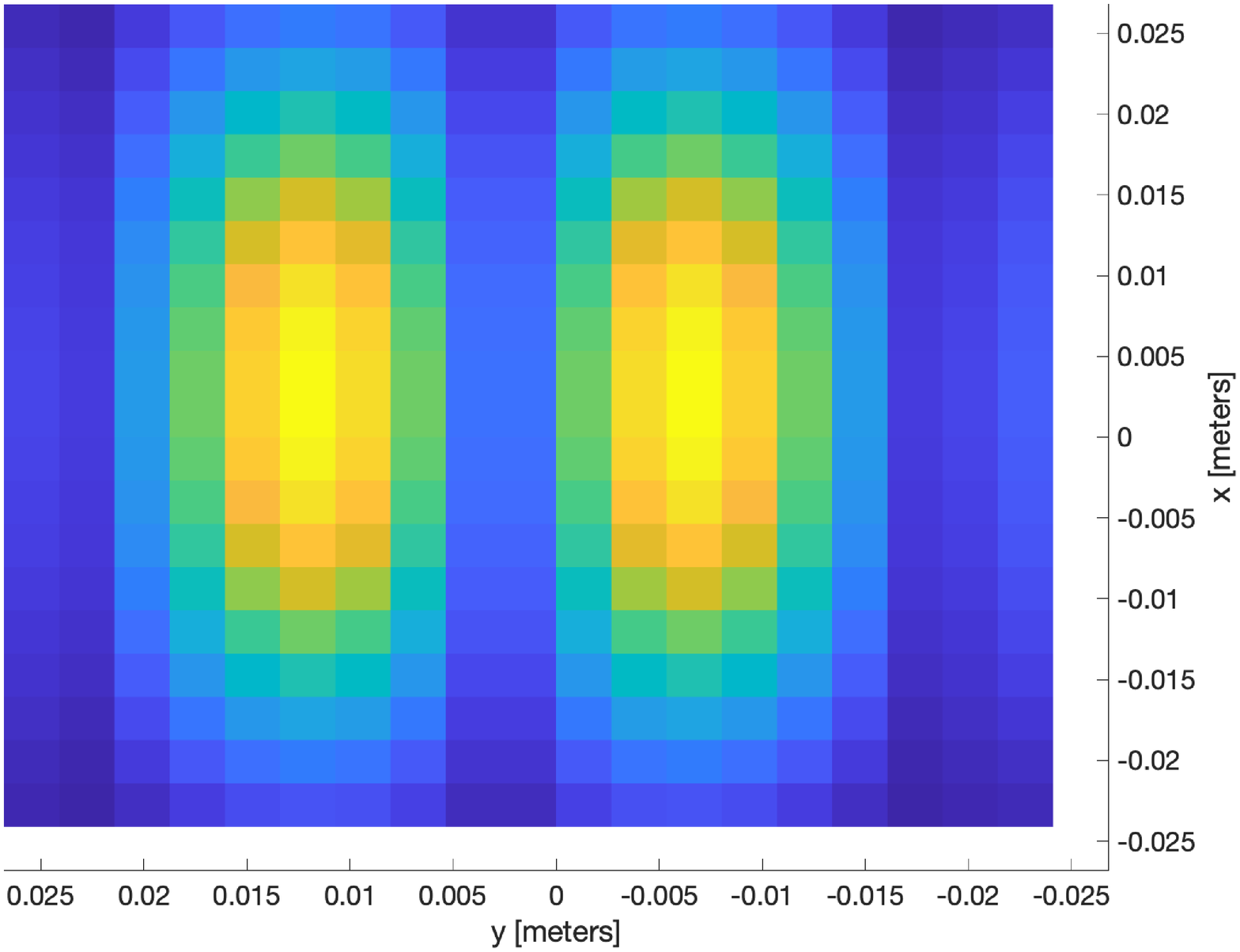}}
\caption{Amplitude of the $x$-component of eigenfunctions $\{\bphi_n(\boldr)\}$ (transmit SIS).}
\label{Fig:eigenfunctions_mag_TX}
\end{figure}

\begin{figure}[t]
\centering
\subfigure[$n=1$]{\includegraphics[clip,width=0.49\columnwidth]{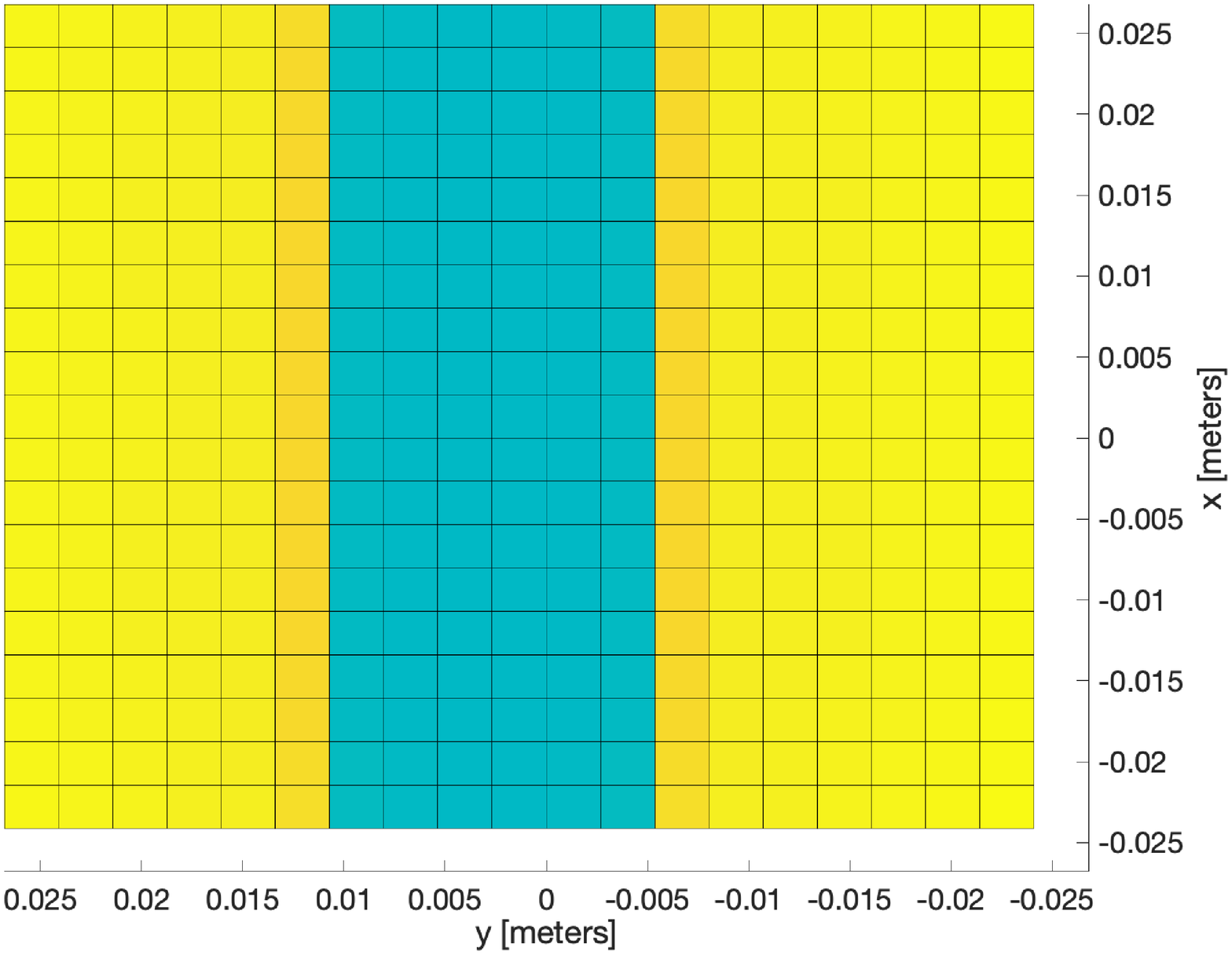}}
\subfigure[$n=2$]{\includegraphics[clip,width=0.49 \columnwidth]{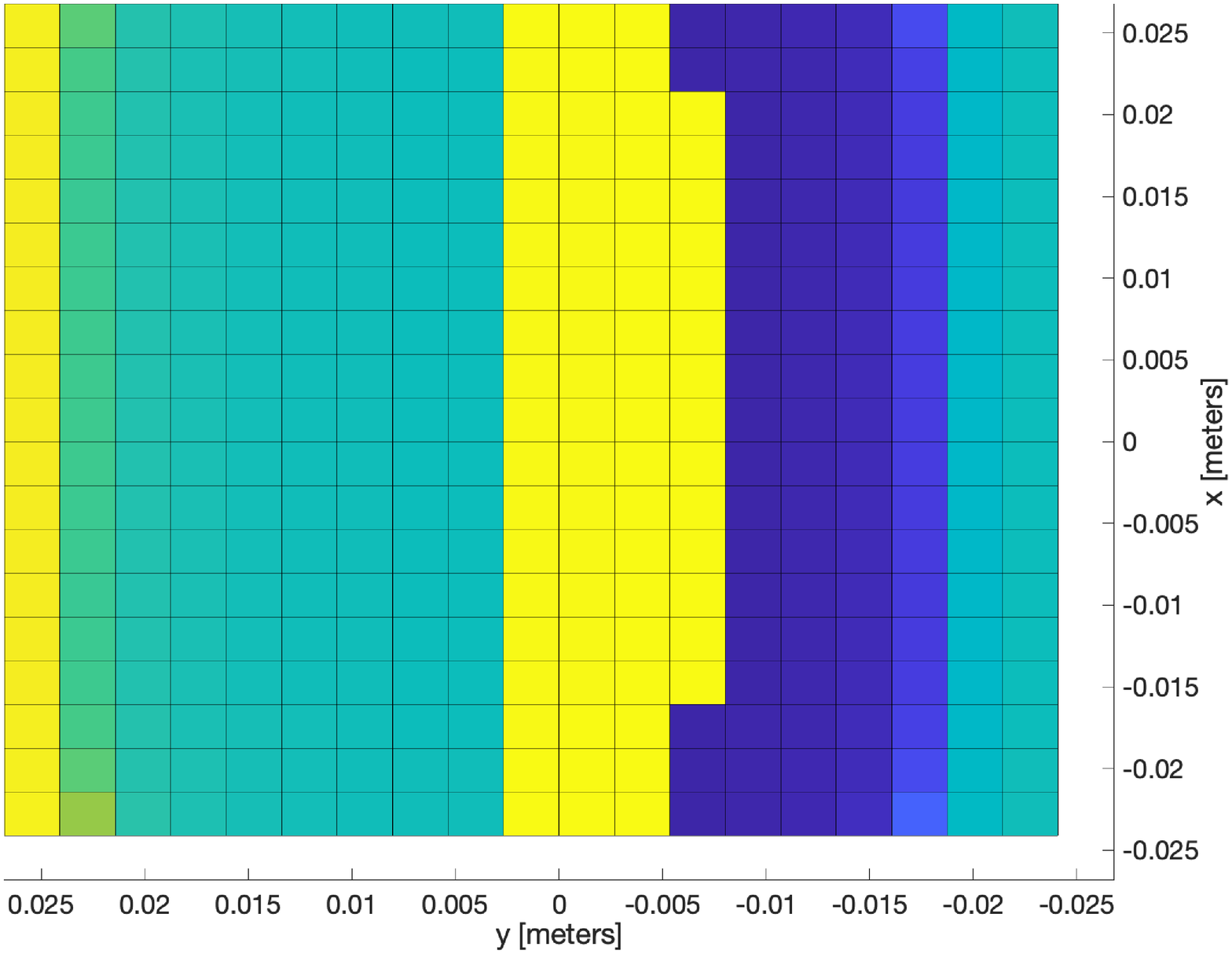}}
\caption{Phase of the $x$-component of eigenfunctions $\{\bphi_n(\boldr)\}$ (transmit SIS).}
\label{Fig:eigenfunctions_phase_TX}
\end{figure}

To get a qualitative idea about the shape of the corresponding eigenfunctions, 
in Figs.  \ref{Fig:eigenfunctions_mag}-\ref{Fig:eigenfunctions_phase_TX} the amplitude and phase of the $x$-component of eigenfunctions $\{ \bpsi_n(\boldr)\}$, for $n=1,2,\ldots 4$, and $\{\bphi_n(\boldr)\}$, for $n=1,2$, are reported, respectively,  under the same parameters used for the results in Fig. \ref{Fig:DoF}.  
For instance, Figs. \ref{Fig:eigenfunctions_mag}a and \ref{Fig:eigenfunctions_phase}a show the electric field observed at the receive \ac{LIS} when the exciting current $\bphi_1(\boldr)$,  corresponding to the largest coupling (i.e., the largest eigenvalue $\xi_1^2$)  reported in Figs. \ref{Fig:eigenfunctions_mag_TX}a and \ref{Fig:eigenfunctions_phase_TX}a,  is considered. 
 From these figures one can notice that orthogonality does not involve in general non-overlapped waves. In fact, the received waves in Figs. \ref{Fig:eigenfunctions_mag}a  and \ref{Fig:eigenfunctions_mag}b (or Figs. \ref{Fig:eigenfunctions_mag}c  and \ref{Fig:eigenfunctions_mag}d) are almost overlapped, but the particular phase distribution deriving from the eigenfunction problems, reported in Fig. \ref{Fig:eigenfunctions_phase}, guarantees the orthogonality between them. 
This means that classical beamforming or focusing schemes aiming at obtaining spatially non-overlapped waves are not in general optimal when using \acp{LIS}. Similar considerations can be done with reference to the transmit \ac{SIS} by observing Figs. \ref{Fig:eigenfunctions_mag_TX}a and \ref{Fig:eigenfunctions_mag_TX}b. Obviously, the generation of such eigenfunctions require a certain level of flexibility in the antenna configuration and signal processing capabilities which implies the adoption of dedicated architectures \cite{Eldar:19}.

\begin{figure}[t]
\centerline{\includegraphics[width=1\columnwidth]{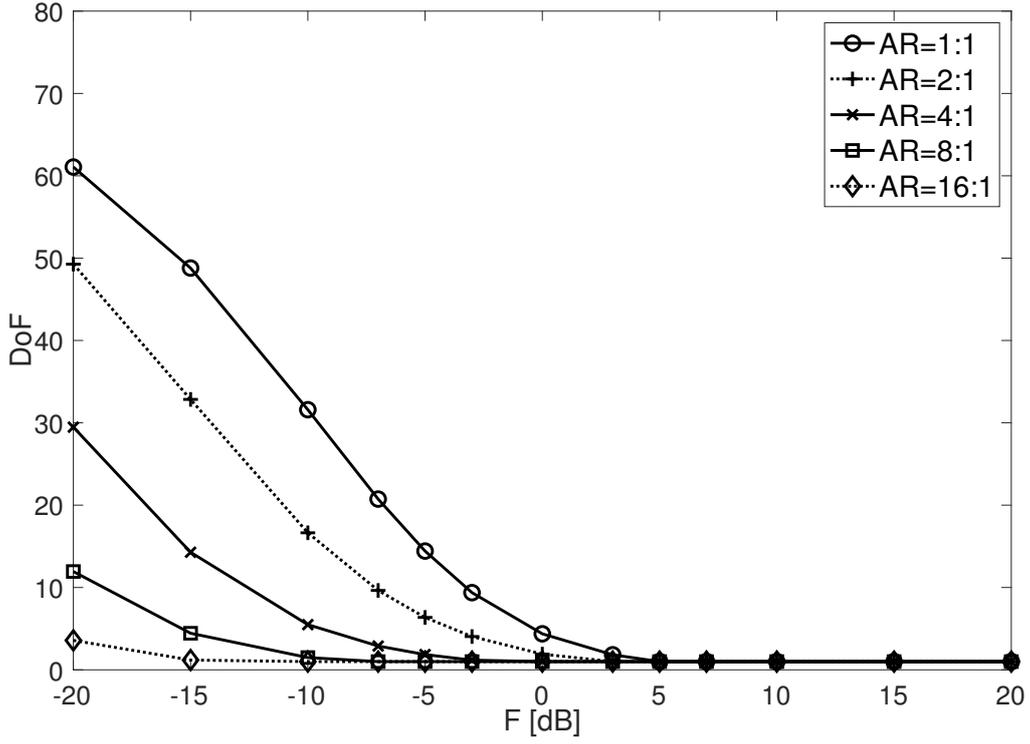}}
\caption{\ac{DoF} vs $F=d^2/\Ar$ for perpendicular surfaces. $\Ar=25\,$cm$^2$, $f_c=28\,$GHz.}
\label{Fig:DoFP}
\end{figure}

The \ac{DoF} for perpendicular surfaces, given by  \eqref{eq:DPerpendicular}, is reported in Fig.  \ref{Fig:DoFP}  as a function of $F$ for different values of $\AR$ under the same conditions as that of Fig. \ref{Fig:DoF}. As it can be noticed, the achievable \ac{DoF} is less than that obtained for parallel surfaces, which represents the best geometric configuration to maximize the \ac{DoF}.
It this case, the result in \cite{Miller:00}, reported in \eqref{eq:Dmiller}, is not applicable because it is not able to capture the \ac{DoF} along the $z$ direction of the \ac{SIS}. 
 
Interestingly, from  the  results in Figs. \ref{Fig:DoF} and \ref{Fig:DoFP} it turns out that \ac{DoF} significantly larger than 1 can be obtained at practical distances in \ac{LOS} channel condition, which can have important  implications in next generation wireless networks operating at millimeter wave and THz  bands. 
For instance, suppose a typical industrial scenario is considered, where a \ac{LIS} of size $5\times 5\,$m$^2$ is deployed on the factory ceiling at heigh $d=5\,$m.   Supposing the transmitting sensors are equipped with \ac{SIS} of area $\At=25\,$cm$^2$ located close to the floor, from Fig. \ref{Fig:DoF} it follows that the  \ac{DoF}   is $D\simeq 20$ ($F=0\,$dB, $\AR=1:1$). This corresponds to a significant increase of link capacity with respect to the situation where only beamforming gain is exploited and $D=1$. For instance, using \eqref{eq:capacitygain}, the capacity gain at $\SNR=20\,$dB is about $7.76$.
 
This result can be interpreted also from another point of view: in fact, equivalently up to $D/\At \simeq 8,000$ orthogonal links per square meter 
 can be activated, which is very promising for the factories of the future where extremely high nodes densities are expected. 
 In addition, the ability to create wireless links orthogonal at \ac{e.m.} level,  simplifies the channel multiple access, thus significantly reducing the communication latency.

\section{Conclusion}
\label{Sec:Conclusion}

We have shown that  the optimal communication between \ac{LIS}/\ac{SIS} can be formulated as an eigenfunctions problem starting from \ac{e.m.} arguments. 
To obtain high-level descriptions of \ac{LIS}-based communication and to avoid extensive and sometimes prohibitive \ac{e.m.}-level simulations, simple but accurate analytical expressions for the link gain and the  communication modes (i.e., \ac{DoF})  between the transmitter and the receiver  have been derived. 
The obtained expressions  allow to get important insights about the communication between intelligent surfaces and can serve as design guidelines in future wireless networks employing \acp{LIS}.

In particular, it has been shown that the achievable \ac{DoF} and gain offered by the \ac{LIS}-enabled wireless link are determined only by geometric factors normalized to the wavelength, and that the classical Friis' formula is no longer valid in this scenario. The fundamental limits for very large intelligent surfaces have been found to be dependent only on the normalized area of the smallest antenna involved in the communication.  

Another  important result is that using \acp{LIS} one can  exploit the spatial multiplexing even in \ac{LOS} channel condition at practical distances, contrarily to conventional \ac{MIMO} systems that can only exploit \ac{SNR} enhancement (beamforming) when in strong \ac{LOS}. 
 This  opens the possibility to satisfy the challenging requirements of next generation wireless networks  operating at millimeter waves or THz bands  in terms of massive communications and high capacity per square meter. 

Obviously, several practical open issues need to be addressed before such limits can be approached by real systems.  
 For instance,  one fundamental research direction is the design of holographic metasurface technologies capable of  approximating the eigenfunctions  required to reach the fundamental limits with affordable complexity.
 Another issue, which deserves particular attention, is the definition of the regulatory power  emission masks for \acp{LIS}. In fact, the question is whether to define the emission masks at the whole antenna level, as done in current regulations with conventional antennas,  or to define ad hoc emission masks, for instance, related to the effective radiated power (ERP) per square meter (ERP spatial density).

\section*{Acknowledgment}
This paper has received funding from the ATTRACT project funded by the European Commission under Grant Agreement 777222.

\section*{Appendix A}

In this Appendix, we show that 
\begin{align} \label{eq:sumrule}
\sum_n \xi_n^2=  \int_{\Sr} \int_{\St} ||\Green(\boldr -\bolds)||^2   \, d \boldr \, d \bolds  \, .
\end{align}

From \eqref{eq:BilinearExpansion},  tensor $\Green(\boldr -\bolds)$ allows the bilinear expansion 
\begin{align}
\Green(\boldr -\bolds)=\sum_n \xi_n \, \bpsi_n(\boldr) \otimes \bphi_n^{\dag}(\bolds)  \, .
\end{align}

The $kj$th element  of tensor $\Green(\boldr -\bolds)$ can be written as
\begin{align}
\left \{ \Green(\boldr -\bolds) \right \}_{kj}=\sum_n \xi_n \, \left \{ \bpsi_n(\boldr) \right \}_{k}  \cdot \left \{ \bphi_n^{\dag}(\bolds) \right \}_{j} \, ,
\end{align}
then 
\begin{align}
 & \left | \left \{ \Green(\boldr -\bolds) \right \}_{kj} \right |^2  \\
 & \, \, \, \, \, \, \, \, \, \, = \sum_n \sum_m \xi_n \, \xi_m \, \left \{ \bpsi_n(\boldr) \right \}_{k}  \left \{ \bpsi_m^{\dag}(\boldr) \right \}_{k} \left \{ \bphi_n^{\dag}(\bolds) \right \}_{j}   \left \{ \bphi_m(\bolds) \right \}_{j} \, .  \nonumber 
\end{align}

For each $k$ it is 
\begin{align}
&\sum_{j=1}^3 \left | \left \{ \Green(\boldr -\bolds) \right \}_{kj} \right |^2 \nonumber \\
& \, \, \, \, \, \, \, \, \, \,= \sum_n \sum_m \xi_n \, \xi_m \, \left \{ \bpsi_n(\boldr) \right \}_{k}  \left \{ \bpsi_m^{\dag}(\boldr) \right \}_{k}  \sum_{j=1}^3 \left \{ \bphi_n^{\dag}(\bolds) \right \}_{j}   \left \{ \bphi_m(\bolds) \right \}_{j} \nonumber \\
& \, \, \, \, \, \, \, \, \, \, =  \sum_n \sum_m \xi_n \, \xi_m \, \left \{ \bpsi_n(\boldr) \right \}_{k}  \left \{ \bpsi_m^{\dag}(\boldr) \right \}_{k}  \bphi_n^{\dag}(\bolds) \, \bphi_m(\bolds) \, . 
\end{align}

By integrating  in $\St$ with respect to $\bolds$ and thanks to the orthogonality condition \eqref{eq:orthogonal}, we obtain
\begin{align}
 \int_{\St} \sum_{j=1}^3 \left | \left \{ \Green(\boldr -\bolds) \right \}_{kj} \right |^2 \, d\bolds=& \sum_n  \xi_n^2 \, \left \{ \bpsi_n(\boldr) \right \}_{k}  \left \{ \bpsi_n^{\dag}(\boldr) \right \}_{k} \, . 
\end{align}

From the previous result, it follows that
\begin{align} \label{eq:partialresult}
\int_{\St} ||\Green(\boldr -\bolds)||^2 \, d\bolds =& \int_{\St} \sum_{k=1}^3 \sum_{j=1}^3 \left | \left \{ \Green(\boldr -\bolds) \right \}_{kj} \right |^2 \, d\bolds \nonumber \\
=& \sum_n  \xi_n^2 \, \sum_{k=1}^3 \left \{ \bpsi_n(\boldr) \right \}_{k}  \left \{ \bpsi_n^{\dag}(\boldr) \right \}_{k}  \nonumber \\
 = & \sum_n  \xi_n^2 \, \left | \bpsi_n(\boldr) \right |^2 \, .
\end{align}

By integrating \eqref{eq:partialresult} in $\Sr$ and exploiting again the orthogonality condition \eqref{eq:orthogonal}, we obtain the final result \eqref{eq:sumrule}.

\section*{Appendix B}

We show here the derivation of \eqref{eq:DParallel1}  from \eqref{eq:DParallel}.  
Since $\Lx, \Ly \ll d$, setting $\xo=\yo=0$,  \eqref{eq:DParallel} can be expanded as 
\begin{align}
D & \simeq  \frac{k_0^2}{8 \pi ^2} \int_{-\Sx/2}^{\Sx/2} \int_{-\Sy/2}^{\Sy/2} \nonumber \\
&  -\frac{\Umx \, \Upy}{\sqrt{\left(\left(\Umx \right)^2+\Upy \right)^2+d^2} \sqrt{\left(\Upx \right)^2  +\left(\Upy \right)^2+d^2}} \nonumber 
 \end{align}
 \begin{align}
 & +\frac{\Upx \, \Upy}{\sqrt{\left(\Upx \right)^2+\left(\Upy \right)^2+d^2} \sqrt{\left(\Umx \right)^2+\left(\Upy \right)^2+d^2}} \nonumber 
 \end{align}
 \begin{align}
 & -\frac{\Umx \,  \Upy}{\sqrt{\left(\Umx \right)^2+\left(\Umy \right)^2+d^2} \sqrt{\left(\Umx \right)^2+ \left(\Upy \right)^2+d^2}} \nonumber 
 \end{align}
 \begin{align}
 & +\frac{\Upx \, \Upy }{\sqrt{\left(\Upx \right)^2+\left(\Upy \right)^2+d^2} \sqrt{\left(\Upx \right)^2+\left(\Umy \right)^2+d^2}} \nonumber 
 \end{align}
 \begin{align}
 & -\frac{\Upx \,  \Umy }{\sqrt{\left(\Upx \right)^2+\left(\Upy \right)^2+d^2} \sqrt{\left(\Upx \right)^2+\left(\Umy \right)^2+d^2}}  \nonumber
  \end{align}
 \begin{align}
 & +\frac{\Umx \, \Umy}{\sqrt{\left(\Upx \right)^2+\left(\Umy \right)^2+d^2} \sqrt{\left(\Umx \right)^2+\left(\Umy \right)^2+d^2}}  \nonumber 
 \end{align}
 \begin{align}
 & +\frac{\Umx \, \Umy }{\sqrt{\left(\Umx \right)^2+\left(\Upy \right)^2+d^2} \sqrt{\left(\Umx \right)^2+\left(\Umy \right)^2+d^2}} \nonumber 
 \end{align}
 \begin{align}
 & -\frac{\Upx \, \Umy\, d \rx \, d\ry  }{\sqrt{\left(\Upx \right)^2+\left(\Umy \right)^2+d^2} \sqrt{\left(\Umx \right)^2+\left(\Umy \right)^2+d^2}} \,  
  \, ,
 \label{eq:integrand}
\end{align}
where $\Upx=r_x+L_x/2$, $\Umx=r_x-L_x/2$, $\Upy=r_y+L_y/2$, and $\Umy=r_y-L_y/2$.

The integrand of \eqref{eq:integrand} can be approximated with the first-order Taylor double series expansion in $\Lx$ and $\Ly$
\begin{align}
\frac{2 d^2 \Lx \Ly}{\left(d^2+\rx^2+\ry^2\right)^2}+O\left(\Lx^2\right)+ O\left(\Ly^2\right) \, , 
\end{align}
resulting in
\begin{align} 
 D \simeq  &  \frac{  d^2 \Lx \Ly}{\lambda^2}  \int_{-\Sx/2}^{\Sx/2} \int_{-\Sy/2}^{\Sy/2} \frac{1 }{\left(d^2+\rx^2+\ry^2\right)^2} \, d \rx \, d\ry \, ,
\end{align}
which admits a closed-form solution given by \eqref{eq:DParallel1}.  Using similar arguments, a closed-form expression can be derived also for the more general case of $\xo,\yo \neq 0$, but it is not reported here due to space constraints and also because no particular insights can be drawn from it.

\ifCLASSOPTIONcaptionsoff
\fi
\bibliographystyle{IEEEtran}
\bibliography{IEEEabrv,MassiveMIMO,MetaSurfaces,WINS-Books}
\end{document}